\documentclass[sigconf, nonacm]{acmart}
\usepackage{caption}
\usepackage{enumitem}
\usepackage{makecell}
\usepackage{subfigure}
\usepackage{array}
\usepackage{tikz}
\usetikzlibrary{shapes, arrows, calc, positioning}
\usepackage{calc}
\usepackage{algorithm}
\usepackage{algpseudocode}
\usepackage{color,soul}
\usepackage{tcolorbox}
\usepackage{cleveref}
\usepackage{balance}
\usepackage{soul}

\makeatletter
\newcommand\footnoteref[1]{\protected@xdef\@thefnmark{\ref{#1}}\@footnotemark}
\makeatother

\begin{document}
\title{There is No Such Thing as an ``Index''!\\
or:\\ The next 500 Indexing Papers}

\newcounter{savecntr}
\newcounter{restorecntr}
\author{Jens Dittrich}
\affiliation{%
  \institution{Saarland Informatics Campus}
}
\email{jens.dittrich@bigdata.uni-saarland.de}

\author{Joris Nix}
\authornote{Both authors contributed equally to this work.}
\affiliation{%
  \institution{Saarland Informatics Campus}
}
\email{joris.nix@bigdata.uni-saarland.de}

\author{Christian Sch\"on}
\authornotemark[1]
\affiliation{%
  \institution{Saarland Informatics Campus}
}
\email{christian.schoen@uni-saarland.de}

\begin{abstract}
Index structures are a building block of query processing and computer science in general.
Since the dawn of computer technology there have been index structures.
And since then, a myriad of index structures are being invented and published each and every year. 

In this paper we argue that the very idea of ``inventing an index'' is a misleading concept in the first place. It is the analogue of ``inventing a physical query plan''.
This paper is a paradigm shift in which we propose to drop the idea to handcraft index structures (as done for binary search trees over B-trees to any form of learned index) altogether. 
We present a new automatic index breeding framework coined \textit{Genetic Generic Generation of Index Structures~(GENE)}.
$\;$ It is based on the observation that almost all index structures are assembled along three principal dimensions: (1)~structural building blocks, e.g., a B-tree is assembled from two different structural node types (inner and leaf nodes),  (2)~a couple of invariants, e.g., for a B-tree all paths have the same length, 
and (3)~decisions on the internal layout of nodes (row or column layout, etc.).
We propose a generic indexing framework that can mimic many existing index structures along those dimensions.
Based on that framework we propose a generic genetic index generation algorithm that, given a workload and an optimization goal, can automatically assemble and mutate, in other words `breed' new index structure `species'.
 In our experiments we follow multiple goals. We reexamine some good old wisdom from database technology.
Given a specific workload, will GENE even breed an index that is equivalent to what our textbooks and papers currently recommend for such a workload? Or can we do even more?
Our initial results strongly indicate that generated indexes are the next step in designing index structures.
\end{abstract}

\maketitle
\pagestyle{plain} 


\section{Introduction}
\label{sec:intro}

\subsection{Problem~1: Indexes are considered monolithic entities} 
When we database researchers talk about indexes, we use the term \textit{index} like referring to an entity of its own. But is that the case? Let's look at our good old B-tree:
A \textit{B-tree index} consists of two different node types: \textit{inner nodes} and \textit{leaves}. Inner nodes keep 
pointers to other nodes. 
The main purpose of an inner node is to route incoming lookups to other nodes.
In addition, a B-tree index algorithmically preserves a couple of invariants, e.g.~all paths from the root to a leaf have the same lengths, each node only has one parent node (i.e.~nodes are structurally organized into a tree), and so forth.
In addition, all nodes keep data in a specific layout (row or column layout, cache-and SIMD-efficient layouts, etc.) and define which search algorithm to use inside a node (binary search, interpolation, prediction, etc.). Since the publication of the original B-tree paper~\cite{DBLP:journals/acta/BayerM72} almost 50~years ago, the physical organization of B-trees has been improved in a zillion different ways, e.g.~\cite{DBLP:conf/vldb/RaoR99,DBLP:conf/sigmod/RaoR00,DBLP:conf/damon/SchlegelGL09,DBLP:conf/sigmod/KimCSSNKLBD10}. 

But what concretely is \textbf{the entity} ``the index'' in here? So far we only defined two different node types pointing to each other, we added a couple of constraints (fan-outs, tree-structure, concrete physical organization of inner nodes and leaves). We may also add heuristics for invariant maintenance (split and merge). 
But, if we change any aspect of this, do we receive a completely different index? When is it just a variant of an \textit{existing} index? And  when is it a \textit{new} index?
For instance, if we change constraints to allow nodes to have more than one parent, would that be a completely different index entity?
Or is it just that one constraint that changes (with possible implications to other features of the index)?

In this paper, we will introduce the idea of  logical and physical indexes. We will show that most existing indexes can be expressed as a specific \textit{configuration} in a generic logical and physical indexing framework\footnote{Note that we will not introduce this as a software framework as done in~\cite{DBLP:conf/vldb/HellersteinNP95,DBLP:conf/vldb/SeegerBBKSDS01} but rather on a conceptual level.} including B-trees, radix-trees, learned indexes, and even extendible hashing. And those configurations can be combined almost arbitrarily \textit{within the same configuration}.
This opens the book for a myriad of hybrid ``indexes''. For instance, in our framework, one extreme of an index (say a single hash table) can smoothly be morphed into another extreme (say a B-tree style index with all kinds of different layouts and search algorithms inside its nodes).

\subsection{Problem~2: Two completely different methodologies to solve a similar problem}
It is remarkable that there is quite a divide in databases when it comes to designing efficient components of a database system like index structures as opposed to designing query plans.
For index structures, the historic and state-of-the-art approach is to define some performance goals, reason about complexities, design something on a blackboard, and then implement it.
Like that an index (much like any other system component) has to be designed from scratch and then implemented.
Eventually, we receive a piece of software that then (hopefully) serves the original purpose.
In sharp contrast to this, since the 70s and the seminal Selinger paper~\cite{DBLP:conf/sigmod/SelingerACLP79} database researchers follow a completely different, and rather successful, design path when it comes to designing query plans: we automatically assemble complex plans from logical and physical operators.

So why follow two completely different design approaches if at the core these are similar problems?
Once we are in the position to express an ``index'' as a configuration in a generic logical and physical indexing framework, there is one question left: Why should we configure indexes by hand anyways? Why should we handcraft which node type to use, which node-internal search algorithm to use, which data layout, tree-levels to use, etc.? 

If we have different components of an index which can be interchanged freely, plus options to play with, well, then we have an optimization problem!

For this reason, in this paper, we will propose a genetic algorithm that, given a dataset and workload, will automatically determine a suitable logical and physical index configuration. 

\vspace*{0.2cm}

\vspace*{-0.3cm}
\subsection{Problem Statement}

We summarize the two principal problems discussed above into the following problem statement that we will investigate in this work:

\begin{enumerate}[itemindent=0.45 cm,labelsep=0.1cm,leftmargin=0cm]
 \item\label{howtogeneralize} How can we generalize the most important index structures into a common conceptual indexing framework?
 
\item How can we automatically breed index structures using (\ref{howtogeneralize}).

\end{enumerate}

\vspace*{-0.1cm}
\subsection{Contributions}

In this paper we make the following contributions:
\begin{enumerate}[itemindent=0.45 cm,labelsep=0.1cm,leftmargin=0cm]
	\item We introduce a generic index structure framework that makes a clear difference between a logical and a physical indexing framework. This is inspired by the split into logical and physical operators in relational and physical algebras/operators.
	\item We present a genetic algorithm which allows us to automatically generate (breed) efficient index configurations (aka indexes).
	\item We present an extensive experimental evaluation of our approach demonstrating that we can both rediscover existing, previously handcrafted indexes as well as new types of hybrid indexes.
\end{enumerate}
        
The paper is structured as follows: in Section~\ref{sec:generic-indexing:conceptual}, we introduce our logical generic indexing framework. After that, in Section~\ref{sec:generic-indexing:physical}, we introduce our physical generic indexing framework.
Both serve as the basis for Section~\ref{sec:genetic-index-breeding} where we introduce our index breeding approach.
Section~\ref{sec:rw} contrasts our approach to related work. Section~\ref{sec:eval} presents our experimental evaluation. We will conclude and point out a couple of exciting future research directions in Section~\ref{sec:conclusion}.

\vspace*{-0.2cm}


\newcommand{\todo}[1]{\textcolor{blue}{\textbf{TODO #1}}}
\newcommand{\btree}{$\text{B}^{+}$~Tree }
\newcommand{\btrees}{$\text{B}^{+}$~Trees }

\section{Generic Logical Indexing Framework}
\label{sec:generic-indexing:conceptual}

In this section we introduce our  generic logical indexing framework. The physical indexing framework is explained in Section~\ref{sec:generic-indexing:physical}.

  \setul{0.5ex}{0.3ex}
    \definecolor{Green}{rgb}{0,.7,0}
    \definecolor{Red}{rgb}{1,0.0,0.0}
    \definecolor{Blue}{rgb}{0,0.0,1}
    \definecolor{Purple}{rgb}{0.6,0.0,0.6}
    \definecolor{Black}{rgb}{0,0,0}

Descriptions of index structures tend to mix up logical (\textit{what} is done) and physical aspects (\textit{how} is that achieved).
For instance, consider the following sentence taken from a popular textbook:

\begin{tcolorbox}
``A  \setulcolor{Black}\ul{sorted} \setulcolor{Red}\ul{file}, called the \textit{\setulcolor{Black}\ul{data} \setulcolor{Red}\ul{file}}, is given another \setulcolor{Red}\ul{file}, called the \textit{\setulcolor{Black}\ul{index} \setulcolor{Red}\ul{file}}, consisting of  \setulcolor{Black}\ul{key}-\setulcolor{Red}\ul{pointer}-pairs. A  \setulcolor{Black}\ul{search key K} in the \setulcolor{Black}\ul{index} \setulcolor{Red}\ul{file} is associated with a  \setulcolor{Red}\ul{pointer} to a \setulcolor{Black}\ul{data}-\setulcolor{Red}\ul{file} \setulcolor{Black}\ul{record} that has  \setulcolor{Black}\ul{search key K}".\\

[Section ``13.1 Indexes on Sequential Files'' in ``Database Systems --- The Complete Book'' \cite{GMUW02}]
\end{tcolorbox}
In this sentence the logical aspects of the index (black underlines, e.g.~sorted, key, record) and the physical aspects of the index (red underlines, e.g.~file, pointer) are introduced \textit{at the same time} and thus mix up both aspects in the same explanation. In a way this violates physical data independence of the index structure.
We want to clearly separate the logical and physical aspects of an index.



\noindent\textbf{Basic Definitions.}
Any expression $\sigma_{P}(R)$ where $P$ is a predicate defined on a relational schema $[R]: \{[ A_1:D_1, \ldots, A_n: D_n]\}$, i.e., a function $P:[R]\mapsto \{\text{true,false}\}$, is called a \textit{query} on $R$.
The result of a query is $\sigma_{P}(R)\subseteq R$.
Given $[R]$ with an attribute $A_i$ with a corresponding non-categorical one-dimensional domain $D_i$, and two constants $l, h \in D_i, l\leq h$, $\sigma_{l \leq A_i \leq h}(R)$ is a \textit{range query} on $R$.
It selects all tuples $t=(a_1,..,a_i,..,a_n) \in R$ where $a_i$ is contained in the interval $[l;h]$. A range query with $l=h$ is called a point query.

\definecolor{Cerulean}{rgb}{0.0, 0.38, 0.45}
\definecolor{OliveGreen}{rgb}{0.23, 0.32, 0.08}

\subsection{Logical Nodes and Logical Indexes}




\begin{definition}{Logical Node.} \label{def:logicalnode}
A logical node is a tuple $(\text{p},\text{RI}, \text{DT})$ where:

\begin{enumerate}[itemindent=0.5 cm,labelsep=0.1cm,leftmargin=0cm]
\item $\text{p}: [R] \rightarrow D$ is a \textbf{partitioning function} on the schema $[R]$ of the dataset to index, (p may be undefined), 

\item RI is the \textbf{routing information}. It is a function $RI: D \rightarrow \mathcal{P}(N)$ where $N$ is a set of nodes and $\mathcal{P}(N)$ is the power set of $N$. In other words, each element of $D$ (the target domain of $\text{p}$) is mapped to a subset of the nodes in $N$. 
For each outcome of the partitioning function $\text{p}$ we can find a set of associated nodes or the empty set.
Notice that the routing information does neither imply nor assume a specific physical organization including a sort order on its entries (like in B-trees).
RI may be undefined. In the following, we use nodes(RI) for the set of nodes mapped to by RI.

\item DT is the \textbf{data}. It is a set of tuples with relational schema $[R]$, DT may be empty\footnote{In principle, DT could also be defined as a similar function as RI the difference being that RI maps to nodes whereas DT maps to tuples. However, to simplify matters a bit, we stick to a set definition at this point.
Also note that the DT-fields can be used to very naturally support buffer-tree-style indexes~\cite{DBLP:conf/wads/Arge95},  bulkloading mechanisms~\cite{DBLP:conf/vldb/BerckenS01} as well as any form of recursive partitioning algorithm.
}.
\end{enumerate}
\end{definition}

\begin{figure}[h!]
\vspace*{-0.7cm}
\includegraphics[trim = 0mm 320mm 220mm 0mm, clip, width=0.39\textwidth,keepaspectratio,page=3]{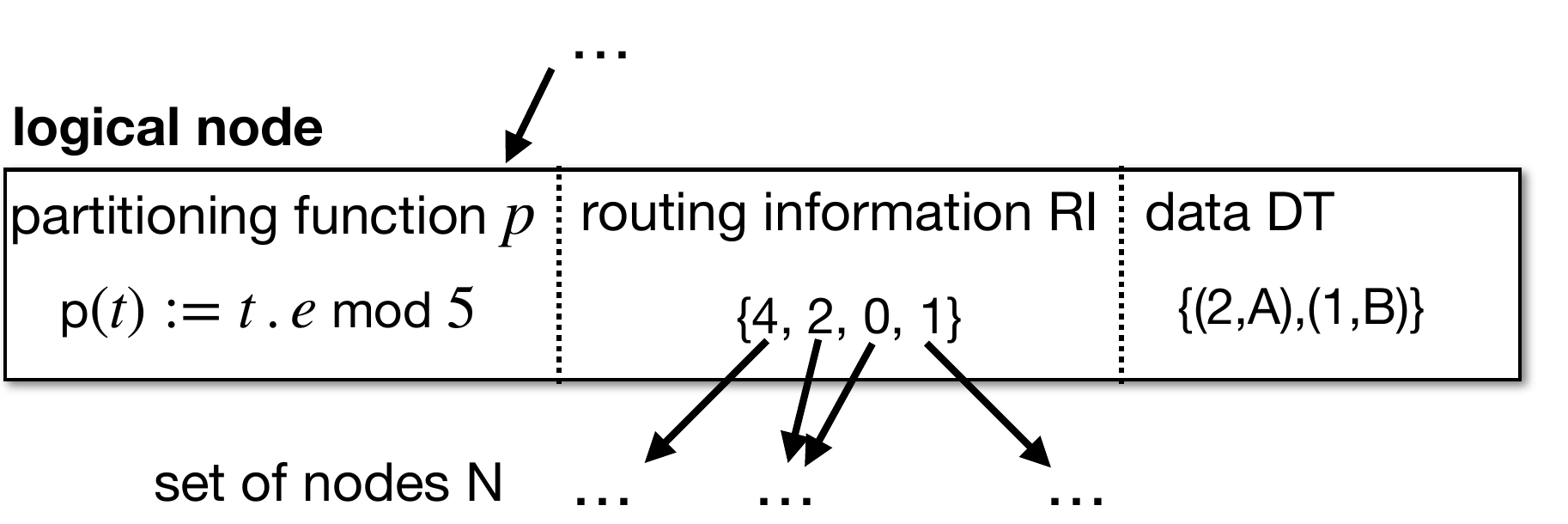}
\vspace*{-0.3cm}
\caption{An example of a logical node with a hash-style partitioning function, four mappings in the routing information RI, and two tuples in the data part.
}
\label{fig:logicalnode}
\vspace*{-0.3cm}
\end{figure}

\begin{figure}[h!]
\subfigure[\textbf{B-tree with ISAM:} Here the partitioning function returns $t.e$. The routing information maps ranges to nodes on the next level. This induces a B-tree-style partitioning. Notice that the common textbook explanation of B-trees showing $k$ pivots and $k+1$ pointers is already a specific physical implementation of this logical index. In addition, this index contains entries on the leaf-level for backward and forward chaining of leaves as in ISAM.
]{
\includegraphics[trim = 0mm 320mm 125mm 0mm, clip, width=0.47\textwidth,keepaspectratio,page=13]{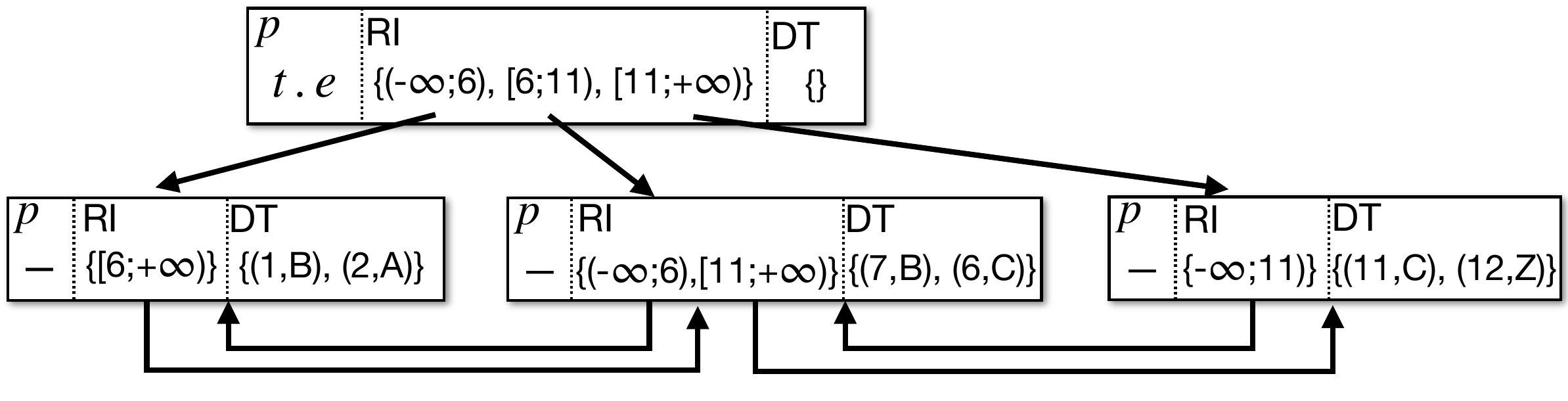}
  \label{fig:logicalindex:example7}
}
\quad 
\subfigure[\textbf{RMI:} Here the partitioning function is a linear function $p(t) = \frac{1}{3} \cdot t.e + 0$ that
squeezes the data into a smaller range ({[}0;12{]} $\rightarrow$ {[}0;4{]}). This is equivalent to a linear regression over the key space. RI groups the data into bins (corresponding to nodes on the next level).
However, $p$ and RI can be set to use any form of regression method and for any node independently.
]{
\includegraphics[trim = 0mm 340mm 110mm 0mm, clip, width=0.472\textwidth,keepaspectratio,page=6]{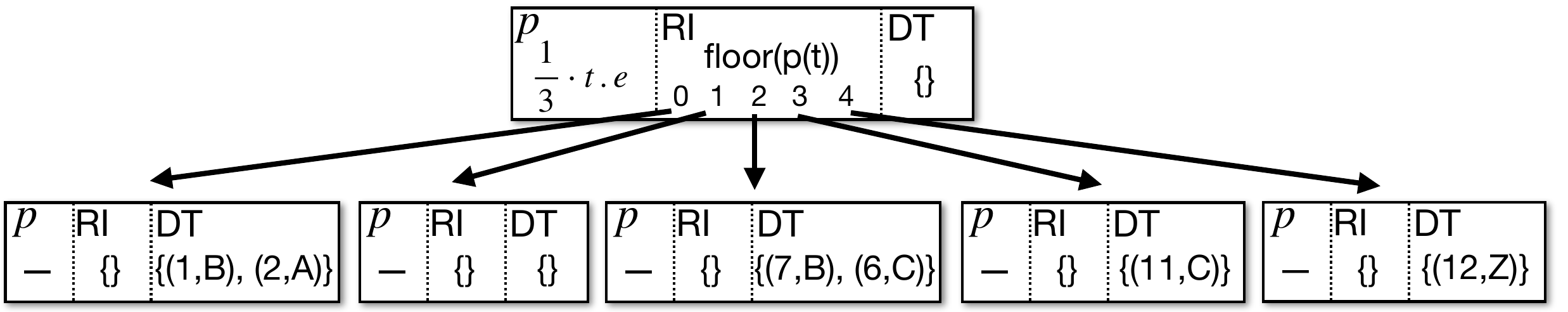}
  \label{fig:logicalindex:example3}
}
\quad 
\subfigure[\textbf{extendible hashing:} Here the partitioning function only considers a suffix of the lowest three bits (\&0x7) of $t.e$. This implies that it partitions exactly like an extendible hashing~\cite{DBLP:journals/tods/FaginNPS79} directory with global depth of three. 
Note that there is no need to create entries for empty `buckets`.
]{
\includegraphics[trim = 0mm 330mm 110mm 0mm, clip, width=0.468\textwidth,keepaspectratio,page=8]{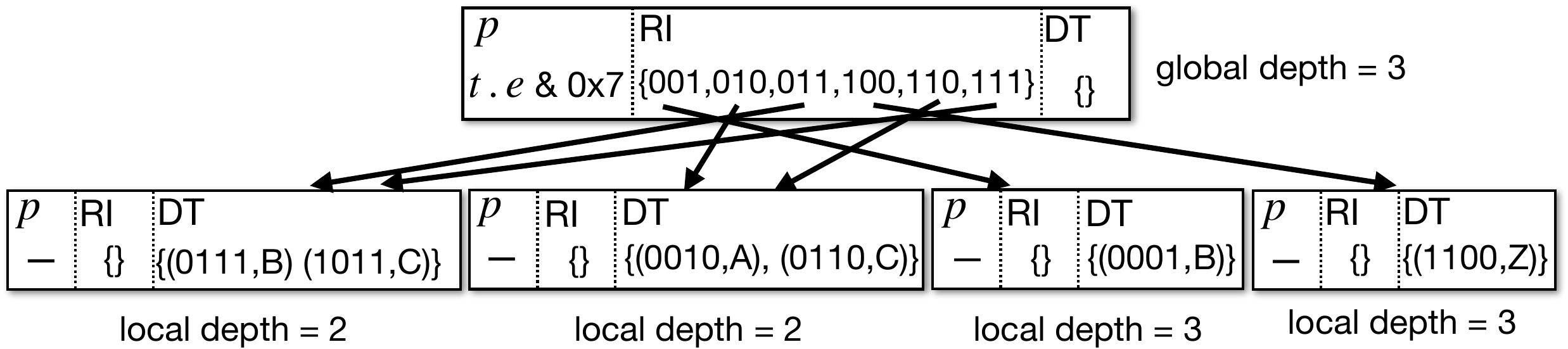}
  \label{fig:logicalindex:example5}
}
\quad 
\subfigure[\textbf{radix tree:} Here the partitioning functions partition the dataset on two adjacent bits each: the root-node partitions on the first two bits of the prefix, the next level on the next two bits. This induces a radix-partitioning. Note that in this example the index is configured to keep at most one tuple per leaf. This can of course be configured. So alternatively, we could force a two-level tree just partitioning on the first two bits. The second level would then keep multiple entries in their DT-fields.]{
\includegraphics[trim = 0mm 280mm 149mm 0mm, clip, width=0.468\textwidth,keepaspectratio,page=12]{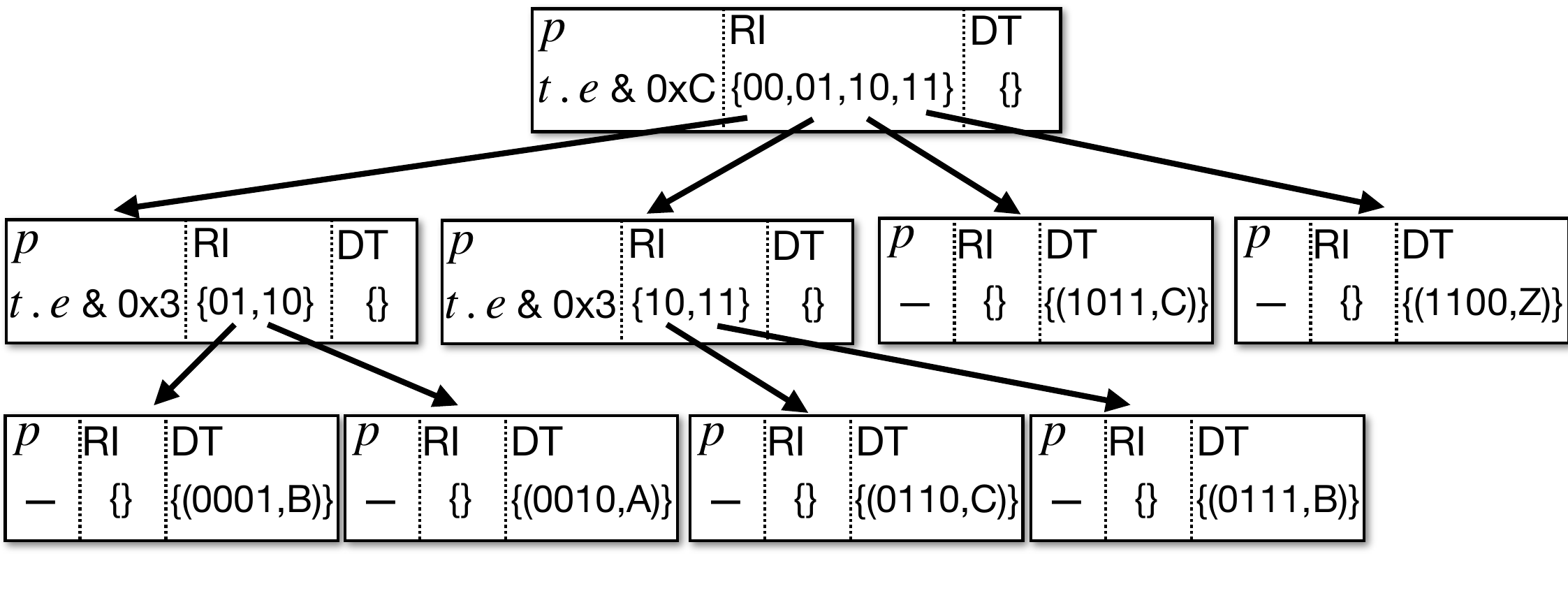}
  \label{fig:logicalindex:example6}
}
\vspace*{-0.4cm}
\caption{The modeling power of our logical indexing framework for traditional indexes. Four special cases of possible logical indexes for the running example. All examples mimic existing and handcrafted (physiological) index structures.
}
\label{fig:logicalindex}
\vspace*{-0.4cm}
\end{figure}

Figure~\ref{fig:logicalnode} visualizes the principal structure of a logical node.
The partitioning function $p$ computes $t.e$ mod 5 which yields a domain $D=\{0,1,2,3,4\}$.
Here, only a subset of $D$ is shown in the visualization of RI, i.e.~3 is not shown as it maps to the empty set. In addition, RI maps 2 and 0 to the same node. 
Moreover, the data part DT contains two tuples $(2,A)$ and $(1,B)$.

\begin{definition}{Complete Logical Index.}
Let $LN$ be a set of logical nodes with $\forall_{n\in LN}: nodes(n.RI) \subseteq LN$.
Then the graph $\lambda=(LN)$ is called a complete logical index.
\label{def:completelogicalindex}
\end{definition}
In other words, only if all routing information in the nodes of $LN$  points to nodes contained in $LN$, we call $LN$ a complete logical index.
At first, this definition sounds  a bit trivial, but this definition makes an important observation that is frequently overlooked: a logical index \textbf{is-a} graph of logical nodes --- and \textbf{nothing else}.

\vspace*{0.2cm}

\begin{figure}[h!]
\includegraphics[trim = 0mm 250mm 0mm 0mm, clip, width=.48\textwidth,keepaspectratio,page=9]{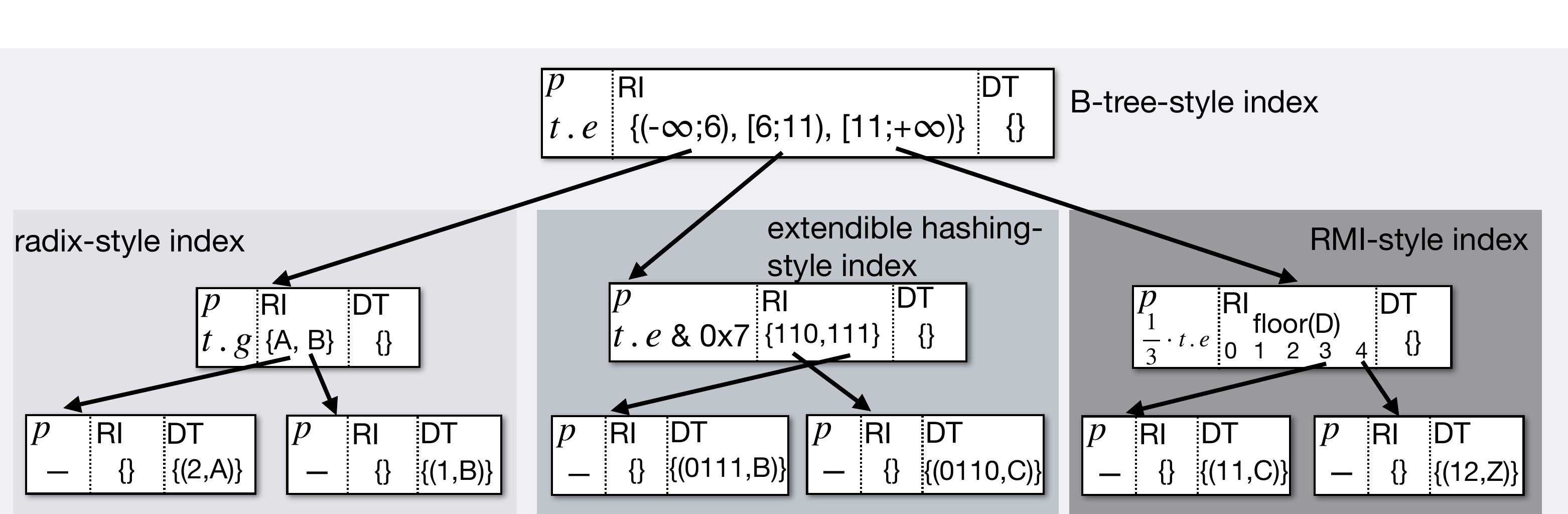}
\vspace*{-0.6cm}
\caption{The modeling power of our logical indexing framework for any form of `hybrid' index. The example combines properties from four different traditional index structures. Notice that this is just one of countless possible examples: any node in this logical index may be exchanged by any other suitable logical node as long as the data in the index is partitioned in a way that all possible queries on the logical index return the correct result set. On this abstraction level it is still undefined \textit{how} data is represented in the different nodes and in particular in the RI-function and the DT-set and \textit{how} we search.
}
\label{fig:logicalindex:hybrid:example1}
\vspace{-0.2cm}
\end{figure}

\noindent\textbf{Running Example.} Figure~\ref{fig:logicalindex} illustrates the modeling power of our framework and shows \textit{four possible} logical indexes for a running example $[R]= \{[e:\text{int}, g:\text{char}]\}$. $R=\{ {(2,\text{A})},{(7,\text{B})},{(1,\text{B})}, {(6,\text{C})},$ ${(12,\text{Z})}, {(11,\text{C})}\}$.
Notice that in all these examples the DT-fields are empty for internal nodes. The implications of allowing data in internal nodes are however considered future work and will therefore not be investigated in this paper.
Figure~\ref{fig:logicalindex:hybrid:example1} demonstrates how we can model arbitrary `hybrid' logical indexes.

\subsection{Logical Queries}




\begin{definition}{\textit{RQ: Result of a Range Query on a Logical Index.}}
	Given a range query with predicate $P:=l \leq A_i \leq h$, a logical index $\lambda$ build upon a relation $R$ and a non-empty start node-set $SN \subseteq LN$, the result set of the range query is given by:
	\[
	\text{RQ}(P,SN) := \bigcup_{n\in SN} \bigg(\underbrace{\sigma_{P}(n.DT)}_{\text{data in }n} \cup \;\text{RQ}\Big(\;P,\bigcup_{t \in R, l\leq t.A_i \leq h} n.\text{RI}\big(n.p(t)\big)\Big)\bigg)
	\] 
	\label{def:RQ:homogeneous}
	\vspace*{-0.3cm}
\end{definition}



\begin{figure}[th!]
\includegraphics[trim = 0mm 255mm 179mm 0mm, clip, width=.48\textwidth,keepaspectratio,page=4]{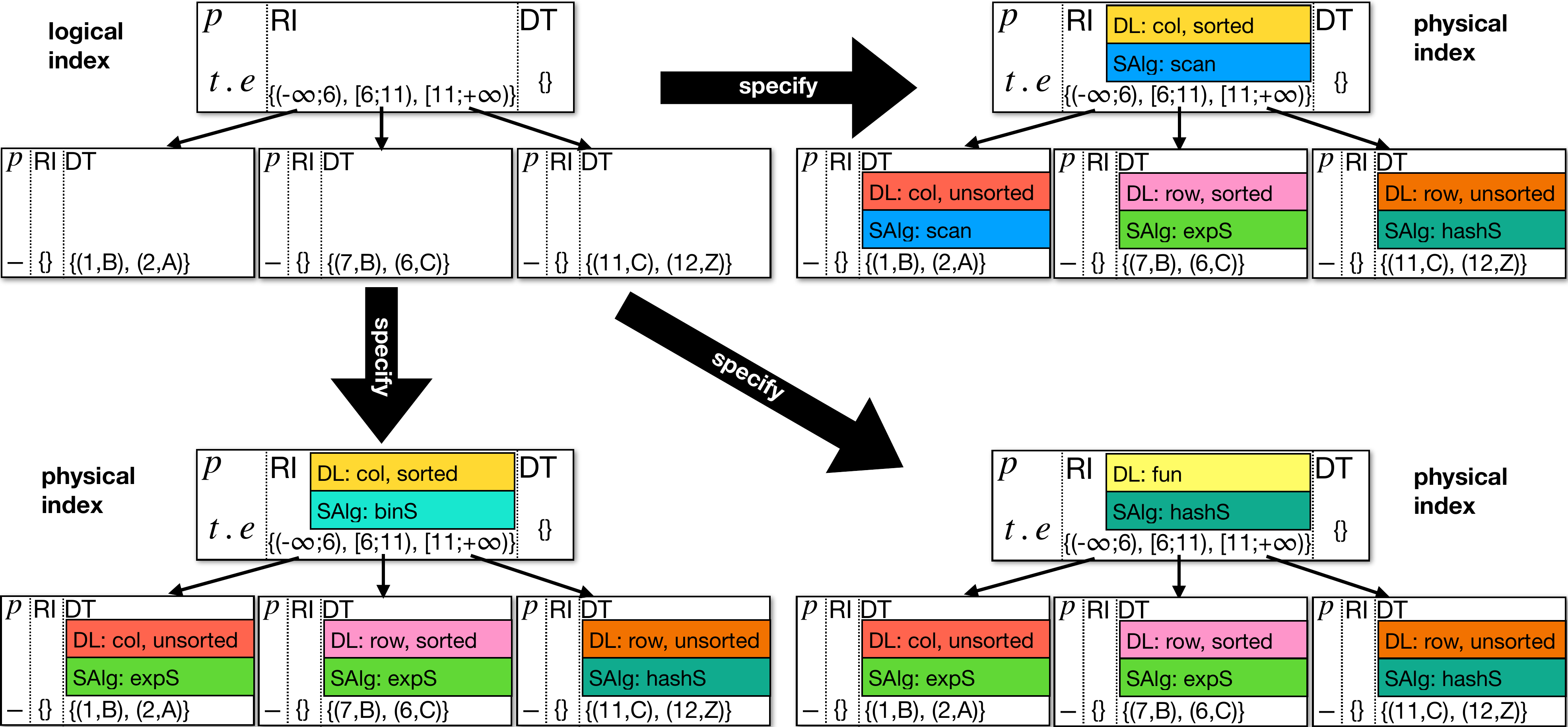}
\vspace*{-0.5cm}
\caption{The arrows show some possible transitions from a logical to a physical index (we specify an algorithm and/or a data layout). Notice that neither the partitioning tree nor the assignment of data to nodes are changed in this process.
}
\label{fig:logica2physicalindex:example1}
\vspace*{-0.5cm}
\end{figure}



Notice that the set semantics will implicitly remove duplicates which in a physical graph-structured index (possibly not obeying set semantics) may result from visiting nodes multiple times.

Also note that this query will recursively traverse the graph for all qualifying nodes in the RI-fields.
This is fine for a strictly tree-structured index, however, as soon as we do not have a tree-structure anymore but a more general DAG, it may become possible that, given a set of start nodes $SN$, certain nodes are reachable via \textit{multiple} paths.
For a general graph, the implementing algorithm has to be modified to not visit nodes multiple times.
%
%
\begin{definition}{\textit{Correctness of a Logical Index.}} \label{def:correctlogicalindex}
Let $\lambda=(LN)$ be a complete logical index. Let $SN$ be an arbitrary non-empty subset of start nodes: $SN \subseteq LN$. Let $DT_{\lambda:}=\bigcup_{n\in LN} n.DT$ be the data contained in $\lambda$.  Let $\sigma_{P:=l \leq A_i \leq h}(R)$ be a \textit{range query} on $R$.
If\[
\forall_{l,h}: \sigma_{l \leq A_i \leq h}(DT_{\lambda}) = \text{RQ}(P,SN),
\]
then $\lambda$ is called a correct logical index w.r.t~$SN$.
\end{definition}
Notice that the correctness of an index depends on whether data is placed into the different DT-sets according to the properties of the different partitioning functions used at the various nodes.
Furthermore, the start nodes $SN$ must be chosen such that all qualifying data can be reached by the range query.
For instance, in a tree-structured index picking the start node is trivial: we call it `the root node'. In a general graph structure, which may even be disconnected, things can become more complex, i.e.~we might have multiple `root nodes', i.e.~all nodes that cannot be reached from any other node of the index, or even no root nodes (in case of a cyclic graph). This discussion is beyond the scope of this paper and therefore in the following, we will only consider correct, DAG-structured indexes and assume that $SN$ is chosen accordingly.

\section{Generic Physical Indexing Framework}
\label{sec:generic-indexing:physical}

As we just have defined logical indexes (our counterparts to the logical relational algebra operators), now, we can proceed to devise physical indexes (our counterparts to physical operators). 


\textit{For each logical node}, \textbf{and} \textit{for each of its RI and DT-part}
we eventually have to specify \textit{how} to realize it.  We do this by making a \textit{physical} decision on the search algorithm (Section~\ref{gindexing:algorithms}) and the data layout to use for that set (Section~\ref{gindexing:datalayouts}). Or, we delegate those decisions by using a nested index (Section~\ref{gindexing:nesting}).

Any index where for all its nodes the data layouts and algorithms are sufficiently specified, is called a \textit{physical index}.

%

\subsection{Specify Search Algorithm}
\label{gindexing:algorithms}

We decide which search algorithm to use for searching (key/value)-pairs in RI and/or DT. 
Note that all search algorithms stop once a qualifying key was found, i.e.~we found the corresponding entry in RI or we have an exact key match in DT. 
The principal options are as follows:
\noindent\textbf{(1)~scan:} linear search through all entries, for each key check if it qualifies,
\noindent\textbf{(2)~binS:} binary search 
\noindent\textbf{(3)~intS:} interpolation search, iteratively compute slope and intercept, i.e.~a linear function, for \emph{left} and \emph{right} key, predict key location \emph{pred} and reduce search area to [left, pred] or (pred, right] respectively until \emph{key} qualifies. 
\noindent\textbf{(4)~expS:}
exponential search, start with the first entry, increase exponent $i$ for key position specified by $2^i$ until \emph{key} is greater than the search value, use binary search (or any other suitable method) inside range [$2^{i-1}$, end].
\noindent\textbf{(5)~hashS:} chained hashing (or any other suitable hashing variant), use the underlying hash function to compute the location of the \emph{key} (and its associated mapping).
\noindent\textbf{(6)~linregS:} linear regression (or any other form of approximation and/or learning), compute slope and intercept, i.e.~linear function, for all data points, compute error bounds, predict key location \emph{pred} and use linear search (or any other suitable error correction method) inside [pred - lower error bound, pred + upper error bound].
\noindent\textbf{(7)~hybridS:} any suitable hybrid algorithm (i.e.~a composite of the former options).


\vspace*{-0.1cm}

\subsection{Specify Data Layout}
\label{gindexing:datalayouts}

We decide which data layout to use for representing the data from RI and/or DT. To define a data layout, we have to specify the following:
\noindent\textbf{(1)~col vs row:}  key/value-pairs are in row or col layout.
\noindent\textbf{(2)~func:} we use a function to specify the RI and/or DT-mapping, thus we do not need to represent pivots and/or data and therefore do not need a data layout. 
As discussed in Definition~\ref{def:logicalnode} already, we assume the DT-fields to be actual sets even though they could be modeled as a more general mapping as well.
\noindent\textbf{(3)~unsorted vs sorted:} the entries are (or are not) sorted by their key.
\noindent\textbf{(4)~comp:} the entries are compressed (and how exactly, i.e.~which compression method).
\noindent\textbf{(5)~hybridDL} any suitable hybrid data layout  (i.e.~any composite of the former options).
Notice that some of these data layout decisions cannot be made independently from the search algorithms to use, e.g.~binary search implies a sorted data layout.
Figure~\ref{fig:logica2physicalindex:example1} shows an example of a logical index that by specifying the search algorithms and data layouts may be transformed into different physical indexes.

\vspace*{-0.1cm}

\subsection{Specify by Nested Logical or Physical Index}
\label{gindexing:nesting}

We make a decision to specify RI and DT by a nested physical index. 
Notice that this is not equivalent to the recursively reachable set of nodes pointed to by one particular RI. Nesting is about representing the key/value-lookup search algorithms and data layout \textbf{inside} a node by another index. For instance, consider a physical binary search tree (BST). If we use such BST to represent and search RI, we basically have a nested physical index in our node. However, this is just a special case, so in theory we can allow for arbitrary nested indexes at this point.


%
%





%
%
%
%
%
%
%
%
%
%
%
%


\section{Genetic Index Breeding}
\label{sec:genetic-index-breeding}
As we just have defined our logical and physical generic indexing frameworks, we proceed to present our genetic algorithm allowing us to automatically generate indexes. 
This is structured as follows:

\begin{enumerate}[itemindent=0.5 cm,labelsep=0.1cm,leftmargin=0cm]
	\item Core algorithm (Section~\ref{sec:corealgorithm}),
	\item Initial population generation (Section~\ref{sec:populationgeneration}),
	\item The set of applicable mutations describing possible changes to individual logical and physical index structures (Section~\ref{subsec:Mutations}), and
	\item The fitness function used to measure the performance of individual physical index structures (Section~\ref{subsec:Fitness}).
\end{enumerate}


\begin{algorithm}
	\begin{footnotesize}
	\algnewcommand{\LineComment}[1]{\State \(\#\) #1}
	\begin{algorithmic}[1] 
		\Function{InitPopulation}{$DS, s_{\text{init}}$}
		\label{func:InitPopulation}
		\State $\Pi = \emptyset$ \Comment{initialize population with empty set}
		\For{$(i=0; i<s_{\text{init}}; i++)$} \Comment{create $s_\text{init}$ initial indexes}	 \label{alg:g3oi:ip:loop} 
		\State $\pi =$ buildAndPopulateRandomIndex$(DS)$  \label{alg:g3oi:ip:bulkload} \Comment{build and populate index}
		\State $\Pi = \Pi \cup \{\pi\}$  \label{alg:g3oi:ip:add} \Comment{add index to population $\Pi$}
		\EndFor
		\State \textbf{return} $\Pi$ \Comment{return population $\Pi$}
		\EndFunction
		\Statex
		\Function{TournamentSelection}{$\Pi, s_{\text{T}},W$}
		\label{func:TournamentSelection}
		\State $T =$ sample\_subset$(\Pi, s_{\text{T}})$  \label{alg:g3oi:ts:sample} \Comment{draw random subset $T \subseteq \Pi$ of size $s_{\text{T}}$}
		\State $\pi_{\text{min}} =$ arg~min$_{\pi \in T} \; f(\pi, W)$ \Comment{select fittest individual $\pi_{\text{min}}$ in $T$ under $W$}\label{alg:g3oi:ts:sbest}
		\State $\tilde{t} = \text{median\_fitness}(T)$\Comment{compute median fitness of all $\pi \in T$} \label{alg:g3oi:ts:median}
		\State \textbf{return} $(\pi_{\text{min}}\;,\; \tilde{t})$ \Comment{return fittest individual $\pi_{\text{min}}$ and median fitness $\tilde{t}$} \label{alg:g3oi:ts:return}
		\EndFunction
		\Statex
		\Function{GeneticSearch}{$g_{\text{max}}, s_{\text{init}}, s_{\text{max}}, s_{\Pi}, s_T, s_{\text{ch}}, DS, MD, ND, W$}
		\label{func:GeneticSearch}
		\State $\Pi = \text{InitPopulation}(s_{\text{init}}, DS)$ \Comment{initialize population} \label{alg:g3oi:gs:initi}
		\For{$(i = 0; i < g_{\text{max}}; i++)$}  \label{alg:g3oi:gs:loopouter}  \Comment{perform $r_\text{max}$ iterations/generations}		
		\State $(\pi_{\text{min}}, \tilde{t}) = \text{TournamentSelection}(\Pi, s_T,W)$ \Comment{run tournament selection} \label{alg:g3oi:gs:ts}
		\For{$(j = 0; j < s_{\text{max}}; j++)$} \label{alg:g3oi:gs:loopinner} \Comment{create $s_\text{max}$ mutations}		 
      
		\State $m=$ draw\_mutation$(MD)$    \Comment{draw from mutation distribution}\label{alg:g3oi:gs:draw}
		\State $n=$ draw\_node$\big(ND(\pi_\text{min}, m)\big)$    \Comment{draw from node distribution}\label{alg:g3oi:gs:node}
		\State $ph=$ draw\_phys$\big(PD(m,n)\big)$    \Comment{draw from phys distribution}\label{alg:g3oi:gs:nodephys}
		\State $\pi_{\text{mut}} = m(\pi_{\text{min}},n,ph)$\Comment{perform  mutation} \label{alg:g3oi:gs:mutate}
		\label{line:Mutation}
		\If{$f(\pi_{\text{mut}},W) \leq \tilde{t}$} \Comment{add $\pi_{\text{mut}}$ to $\Pi$ if fitter than median $\tilde{t}$}\label{alg:g3oi:gs:addifbetter}
		\If{$|\Pi| \geq s_{\Pi}$} \label{alg:g3oi:gs:checkcapactiy} \Comment{if capacity exceeded}
		\label{line:PopulationReduction}
		\State $\Pi = \Pi \;\setminus\; $arg max$_{\pi \in T} \; f(\pi, W)$  \label{alg:g3oi:gs:removeworst} \Comment{remove unfittest individual}
		\EndIf		
		\label{line:MedianComparison}
		\State $\Pi = \Pi \cup \{\pi_{\text{mut}}\}$  \label{alg:g3oi:gs:addmutation} \Comment{add index to population}
		\EndIf
		\EndFor
		\EndFor \label{alg:g3oi:gs:loopouterend}
		\State $\pi_{\text{min}}=$ arg~min$_{\pi \in \Pi} \; f(\pi, W)$\Comment{return fittest individual of final population} \label{alg:g3oi:returnbest}
		\State \textbf{return} $\pi_{\text{min}}$
		\EndFunction
	\end{algorithmic}
	\end{footnotesize}
	\caption{Genetic Search Algorithm of GENE}
	\label{alg:GeneticSearch}
	\vspace*{-0.1cm}
\end{algorithm}

The major challenge with a generic indexing framework presented in Section~\ref{sec:generic-indexing:physical} is the intractable search space.
Therefore, we need an optimization method that can cope with such a huge search space. Notice that an intractable search space does not imply that we cannot find a good solution. In fact, entire research communities work on these kind of problems including: planning, reinforcement learning, and genetic optimization. We decided to design our search algorithm based on genetic optimization.
Genetic optimization algorithms have been developed for more than 40 years~\cite{holland1975adaptation}, but recently gained a lot of attention due to growing computational resources. They allow researchers to effectively explore larger search spaces.
Recent surprising, and not widely-known, results include: \textit{genetic algorithms can rediscover state-of-the-art machine learning algorithms}(!)~\cite{real2020automlzero}. Furthermore, they can devise yet unknown mathematical equations~\cite{cranmer2020discovering}.
Genetic optimization tasks are very domain specific as possible mutations and the performance measure depend heavily on the concrete task. 

%
%
%

\vspace*{-0.2cm}

\subsection{Core Algorithm} \label{sec:corealgorithm}

The general design for our algorithm follows the principal of evolution which is known from nature: 
We start with the main function \textsc{GeneticSearch}~(line~\ref{func:GeneticSearch}).
We start by initializing a \textit{population} of individuals (line~\ref{alg:g3oi:gs:initi}), in our case a set of physical index structures $\Pi := \{\pi | \pi\text{ is a physical index}\}$ (see function \textsc{InitPopulation}, line~\ref{func:InitPopulation}).
To create the initial population, we build and populate $s_{\text{init}}$ physical index structures (line~\ref{alg:g3oi:ip:bulkload}) and add them to the population~$\Pi$ (line~\ref{alg:g3oi:ip:add}).
This build process is described in more detail in Section~\ref{sec:populationgeneration}.
Now, we enter the central iteration: we perform $g_{\text{max}}$ iterations in genetic search  (lines~\ref{alg:g3oi:gs:loopouter}--\ref{alg:g3oi:gs:loopouterend}). We start by tournament selection (line~\ref{alg:g3oi:gs:ts}), see function \textsc{TournamentSelection} (line~\ref{func:TournamentSelection}).
We select a sample of size $s_\text{T}$ of the current population $\Pi$ (line~\ref{alg:g3oi:ts:sample})
from which we select the fittest index $\pi_{\text{min}}$ (line~\ref{alg:g3oi:ts:sbest}). We keep a trace of the fitness of physical indexes to never evaluate indexes multiple times.
We compute the median fitness $\tilde{t}$ of sample $T$ (line~\ref{alg:g3oi:ts:median}) and return both $\pi_{\text{min}}$ and $\tilde{t}$ (line~\ref{alg:g3oi:ts:return}) to the GeneticSearch function (line~\ref{alg:g3oi:gs:ts}).
Then, we enter the mutation loop (line~\ref{alg:g3oi:gs:loopinner}). The core idea is to compute $s_{\text{max}}\geq 1$ mutations for index $\pi_{\text{min}}$.
We draw a random mutation $m$  from a precomputed distribution of mutations $MD$ (line~\ref{alg:g3oi:gs:draw}). For the mutation $m$ we draw a start node $n$ to be used for this mutation (line~\ref{alg:g3oi:gs:node}) as well as a physical implementation $ph$ (line~\ref{alg:g3oi:gs:nodephys}).
The mutations and distributions are described in detail in Section~\ref{subsec:Mutations}. 
%
\begin{table}
\begin{footnotesize}
\begin{tabular}{|l|p{6.5cm}|} \hline
\textbf{Symbol} & \textbf{Meaning} \\ \hline\hline
$\lambda$ & logical index \\
$\pi$ & physical index \\
$\Pi$ & population \\
$s_{\text{init}}$& initial size of the population \\
$s_{\Pi}$& maximum number of indexes in population \\
$g_{\text{max}}$& number of generations \\
$s_{\text{max}}$ & number of mutations created and evaluated in a single iteration\\
$s_{\text{T}}$& size of sample in tournament selection \\
$s_{\text{ch}}$ & maximum length of a mutation chain applied in one iteration\\
$DS$ & dataset \\ 
$\pi_\text{min}$ & best individual in tournament selection\\
$\pi_\text{mut}$ & mutated element\\
$\tilde{t}$ & median fitness \\
$MD$ & probability distribution of mutations \\
$m$ & a single mutation \\
$ND(\pi, m)$ & probability distribution of nodes \\
$PD(m,N)$ & probability distribution of physical implementations \\
$W$ & workload of queries \\
$f(\pi, W)$ & fitness of a physical index\\
\hline
\end{tabular}
\end{footnotesize}
\caption{Symbols used.}
\vspace*{-1cm}
\end{table}
Then, we perform the actual mutation on $\pi_{\text{min}}$ (line~\ref{alg:g3oi:gs:mutate}) and receive $\pi_{\text{mut}}$.  We originally also experimented with applying chains of mutations (lines~\ref{alg:g3oi:gs:draw} and~\ref{alg:g3oi:gs:mutate}) but it did not show any benefits.
We check, whether the mutated index $\pi_{\text{mut}}$ has a better fitness than the median $\tilde{t}$ (line~\ref{alg:g3oi:gs:addifbetter}).  If it has a better fitness, we check if $\Pi$ exceeds its capacity of maximum allowed physical indexes $s_{\Pi}$ (line~\ref{alg:g3oi:gs:checkcapactiy}). If that is the case, we remove the physical index with the worst fitness from $\Pi$ (line~\ref{alg:g3oi:gs:removeworst}). Then we add $\pi_{\text{mut}}$ to the population $\Pi$ (line~\ref{alg:g3oi:gs:addmutation}).
Once the outer loop terminates, we determine the fittest index from $\Pi$ (line~\ref{alg:g3oi:returnbest}) and return it.

%
%

\subsection{Initial Population Generation}
\label{sec:populationgeneration}


What is a good start population $\Pi$ for the genetic algorithm?
In Algorithm~\ref{alg:GeneticSearch}, function InitPopulation (line~\ref{func:InitPopulation}), we need to define an initial population of individual index structures.
There are several possible dimensions to consider. First, we can change the initial number $s_{\text{init}}$ of indexes in $\Pi$. This basically defines how diverse the initial set of indexes may be.
Second, we should determine how to actually build and populate the initial physical index with data from dataset DS (line~\ref{alg:g3oi:ip:bulkload}). 
There are several options:
\begin{enumerate}[itemindent=0.45 cm,labelsep=0.1cm,leftmargin=0cm]
\item \label{sec:init:pop:generation:1} We start with a single physical node that does not contain data, mutate it, and only then insert the actual data.
We experimented with this approach initially but discarded it quickly due to its high training costs.
Thus we do not support it in our algorithm anymore. 


\item \label{sec:init:pop:generation:2} We start with a single physical node containing all data. For data layout/search method we either randomly pick it or we pick one that we believe works well for the given workload. 

\item \label{sec:init:pop:generation:3} We use bottom-up bulkloading with the difference that for all nodes the search algorithms and data layouts are picked randomly.
In our current version we exclude hash nodes for inner nodes as we have not defined a radix-partition search method on this data layout yet. 
We will integrate this in future versions of our optimization framework.
The resulting tree is logically similar to a standard B-Tree, the physical nodes however differ considerably.

\item \label{sec:init:pop:generation:4} We start with a population containing a physical index that resembles a state-of-the-art hand-tuned index, i.e.~we define the logical index (including its partitioning functions) as well as the physical nodes. Then we check whether we can still improve that index through our genetic algorithm.
\end{enumerate}

Notice that for options from (\ref{sec:init:pop:generation:1}) to (\ref{sec:init:pop:generation:4}) increasing, 
we postulate that we take away load from GENE, using it increasingly as a refinement tool: The more we start with something already representing a very efficient (or fit, however fitness is defined) index, the more we expect that only small mutations will be performed by GENE.
At least that is what we would believe. In fact, even if we (non-randomly) specify an initial physical index to start with, recall, that GENE has all degrees of freedom to pick mutations, and may surprise us by taking unexpected turns and make different decisions.

\vspace*{-0.2cm}
\subsection{Mutations and their Distributions}
\label{subsec:Mutations}

\newcommand{\mutscale}{0.12}

\begin{figure*}[h!]
    \subfigure[\textbf{M$_1$ Change node type:} change data layout of RI.]{
    \includegraphics[trim = 0mm 307mm 256mm 0mm, clip, scale=\mutscale, keepaspectratio,page=9]{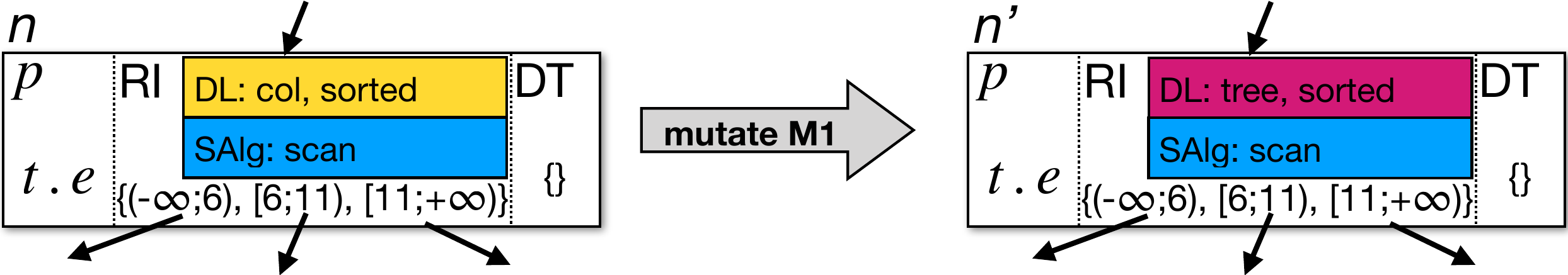}
      \label{fig:m1}
    }
    \quad
    \subfigure[\textbf{M$_2$ Change search method:} change search of~RI.]{
    \includegraphics[trim = 0mm 307mm 256mm 0mm, clip, scale=\mutscale, keepaspectratio,page=10]{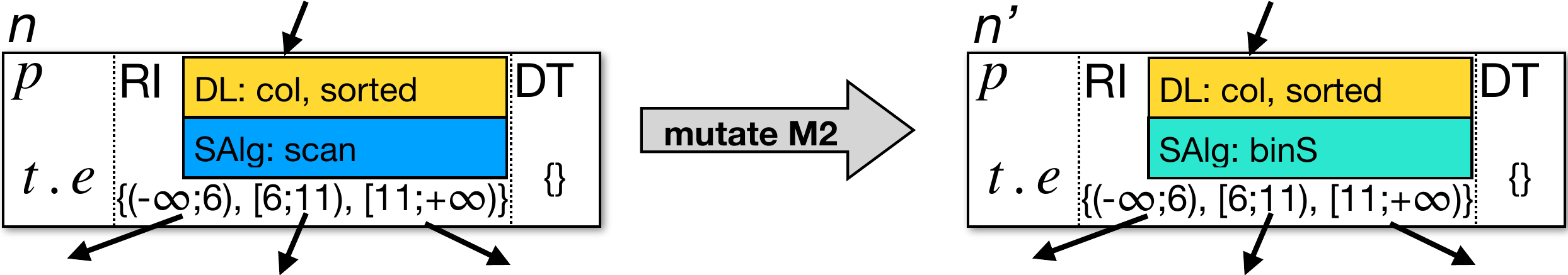}
      \label{fig:m2}
    }
    \quad
    \subfigure[\textbf{M$_3$ Merge nodes horizontally:} merge left \& middle child node.]{
    \includegraphics[trim = 0mm 246mm 116mm 0mm, clip, scale=\mutscale, keepaspectratio,page=11]{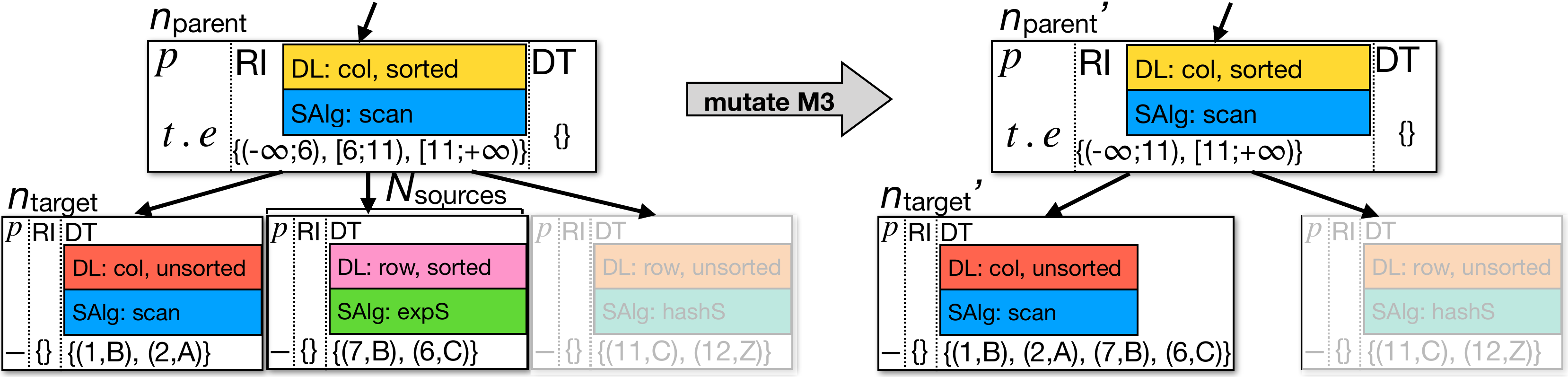}
      \label{fig:m3}
    }
    \quad
    \subfigure[\textbf{M$_4$ Split node horizontally:} split left child node.]{
    \includegraphics[trim = 0mm 247mm 0mm 0mm, clip, scale=\mutscale, keepaspectratio,page=12]{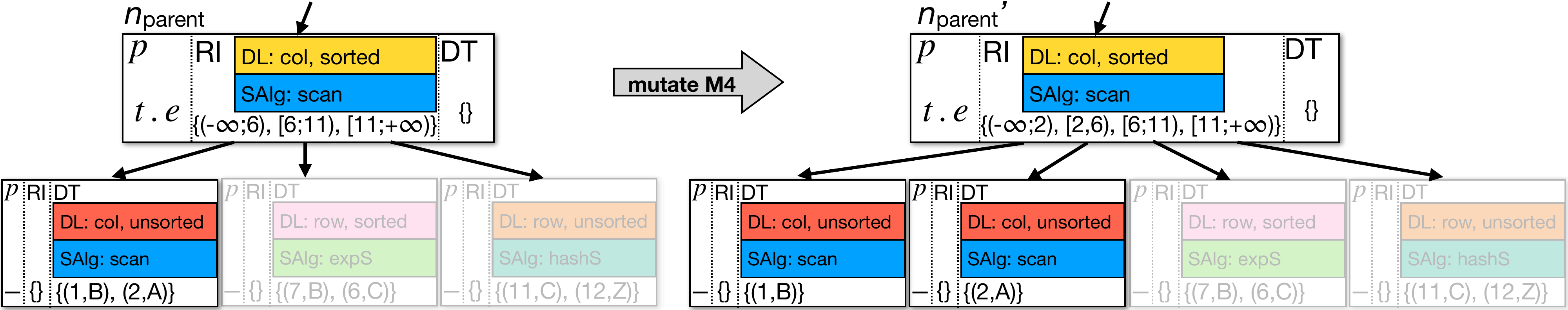}
      \label{fig:m4}
    }
    \quad
    \subfigure[\textbf{M$_5$ Merge nodes vertically:} merge top-level node's left child.]{
    \includegraphics[trim = 0mm 178mm 0mm 0mm, clip, scale=\mutscale, keepaspectratio,page=13]{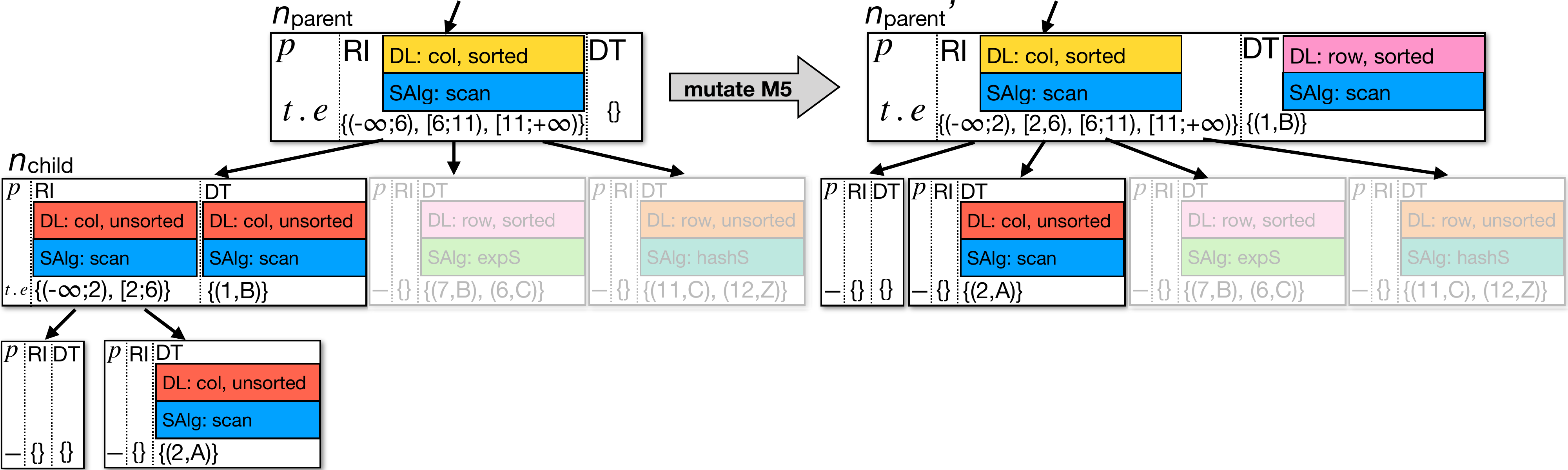}
      \label{fig:m5}
    }
    \quad
    \subfigure[\textbf{M$_6$ Split node vertically:} split left child.]{
    \includegraphics[trim = 0mm 180mm 9mm 0mm, clip, scale=\mutscale, keepaspectratio,page=14]{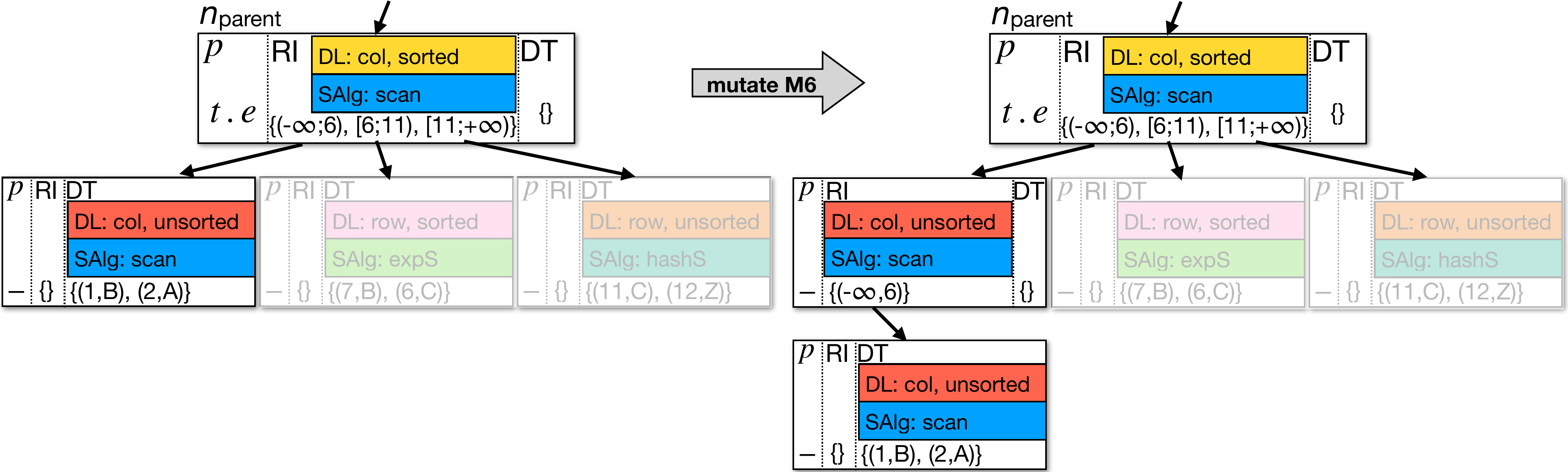}
      \label{fig:m6}
    }
    \quad
    \vspace*{-0.4cm}
    \caption{Performing the mutations described in Section \ref{subsec:Mutations} on actual physical indexes.}
    \label{fig:mutations}
    \vspace*{-0.4cm}
\end{figure*}

In this section we introduce a suitable set of mutations and discuss how they are used in our algorithm.

\noindent\textbf{Mutation.} In our framework, a \textit{mutation} is a function $m:$ Index$ \rightarrow$ Index. 
A mutation takes a single index as input, mutates it, and returns a modified  index.  By `Index' we mean, that either a logical index ($\lambda$) \textit{or} a physical index ($\pi$) is given and a mutated index is returned ($\lambda_\text{mut}$ or $\pi_\text{mut}$).
$\lambda_\text{mut}$ and $\pi_{\text{mut}}$ must preserve the correctness of $\lambda$ and $\pi$. This is inspired by rewrite rules in classical query optimization: there we also only consider rules that are guaranteed to not change the query result.
We will only consider mutations on tree-structured indexes. This is not a restriction of our  generic framework but makes the following mutations a bit more digestible.

\noindent\textbf{Mutation distributions.} We use a probability distribution $MD$ allowing us to assign different probabilities to the different mutations (line~\ref{alg:g3oi:gs:draw}), e.g.~we can give higher probabilities to certain mutations.
Given a mutation $m$ and a physical index $\pi_\text{min}$ we then draw from a second distribution $ND(\pi_\text{min}, m)$ to determine the nodes to use for this mutation (line~\ref{alg:g3oi:gs:node}). Now, we draw from a third distribution $PD(m,N)$ to determine which physical implementation to use for this mutation and node set. 
Setting probabilities to zero within this distribution $PD(m,N)$ excludes combinations of physical data layout and search method which are invalid, e.g. binary search on unsorted data layouts.
Note that these distributions can be created based on microbenchmarks.



\noindent\textbf{Fundamental Mutations.} Our goal is to implement a minimal set of mutations that allow for breeding a huge variety of physical index structures.


\vspace*{-0.1cm}

\begin{enumerate}[label=$\mathbf{M_{\arabic*}}$,itemindent=0.5 cm,labelsep=0.1cm,leftmargin=0cm]
	\item \textbf{Change data layout:} From $n$, we randomly select either its RI- \textit{or} DT-part. 
Then we create a new physical node $n'$ with data layout $n'.dl\neq n.dl$ drawn from $PD(m,N)$ with the same data and routing information as $n$: $n'.DT = n.DT \wedge n'.RI = n.RI$.
The options for data layouts are described in Section~\ref{gindexing:datalayouts}.
If $n$ contains child partitions, we enforce the additional condition $n.dl' \neq \text{hash}$, as our software framework does not (yet) support child partitions in nodes with a hash layout. In $\pi$, we replace $n$ by $n'$.
If $n'.s$ is incompatible with $n'.dl$, we draw a new method from $PD(m,N)$ to ensure correctness.
	Figure~\ref{fig:m1} shows an example:  the input node $n$ has a sorted column-layout. In the index, we replace $n$ by $n'$ which has a tree-layout.
	
	\item \textbf{Change search method:} From $n$, we randomly select either its RI- \textit{or} DT-part. 	Given the existing search method $n.s \in S:= \{\text{scan}, \text{binS}, \text{intS}, \text{expS}, \text{linRegS}, \text{hashS}\} $, we draw an $s' \in S$ with $s' \neq s$ from $PD(m,N)$.
	Then we create a new physical node $n'$ with the new search method $s'$ with the same data and routing information as $n$: $n'.DT = n.DT \wedge n'.RI = n.RI$.
	Figure~\ref{fig:m2} shows an example: the input node $n$ uses a scan as search method. In the index, we replace $n$ by $n'$ which uses binary search.

	\item \textbf{Merge sibling nodes horizontally:}  
	We set node~$n_\text{parent}:=n$ whose RI maps to at least one other node in $\pi$, if not we abort this mutation.
	From the set of nodes mapped to by $n_\text{parent}$ we randomly select a child node  $n_{\text{target}} \in \text{nodes}(n_\text{parent}$.RI$)$. We select a non-empty subset $N_\text{sources} \subseteq  \text{nodes}(n_\text{parent}$.RI$)$  
		of nodes to merge into $n_{\text{target}}$ using the following restrictions: $n_{\text{target}} \notin N_\text{sources} \wedge \forall_{n \in N_\text{sources}}\; n.p = n_{\text{target}}.p$. This implies that the source domain of the routing information function $D$ is equal for all nodes in $N_\text{sources} \cup \{n_{\text{target}}\}$.
We then need to perform updates on two levels of the index: The node $n_{\text{target}}$ that we merge with and the parent node $n_{\text{parent}}$. We start by describing the updates to the node $n_{\text{target}}$.
First we update the data $n_{\text{target}}$.DT  and set it to the union of all data within the merged nodes: 
\[n_{\text{target}}'.\text{DT} = n_{\text{target}}.\text{DT} \;\;\cup \bigcup_{n \in N_\text{sources}} n.\text{DT}.\]
In the following, we also update the routing information function $n_{\text{target}}.\text{RI}$ such that
\[ \forall_{d \in D} n_\text{target}'.RI(d) = n_{\text{target}}.\text{RI}(d) \cup \bigcup_{n \in N_\text{sources}} n.\text{RI}(d) ,\]
where $D$ is the common domain of the RIs in $N_\text{sources} \cup \{n_{\text{target}}\}$.

This ensures that our target node $n_{\text{target}}$ now maps to all child nodes that any node $n\in N_\text{sources}$ previously mapped to, i.e.~we can still reach all child nodes. 
For the parent node $n_{\text{parent}}$ we have to update the routing information $n_{\text{parent}}$.RI such that \[\forall_{d \in n_{D_{\text{parent}}}}\; \forall_{n \in N_\text{sources}}\; n \in n_{\text{parent}}.RI(d)·\]
\[ \Rightarrow n_{\text{parent}}.RI(d) = \{n_{\text{target}}\} \cup n_{\text{parent}}.RI(d) \setminus \{n\}.\]

In other words: We remove all mappings to merged nodes $n\in N_\text{sources}$ and replace them with a new mapping to the node $n_{\text{target}}$.

Notice that the merge operation performed in B-trees is essentially just a specialized version of this general merge mutation.
In a B-tree the number of merged nodes $k$ is typically set to $k=2$ and the nodes must be directly neighboring due to the sorted key domain. 
For our actual implementation, we also restrict ourselves similarly to merges where $|N_\text{sources}| = 1$.
Merge operations with larger source-sets can easily be achieved by recursively executing the merge operation on the same node.
Figure~\ref{fig:m3} shows an example: the set $N_\text{sources}$ contains a single leaf that we want to merge into $n_\text{target}$. To achieve this we first merge all data contained in $N_\text{sources}$.DT into $n'_\text{target}$.DT. As $N_\text{sources}$.RI is empty, we do not have to do anything here. In $n_\text{parent}$.RI, we need to remove the mapping to all nodes in $N_\text{sources}$, in this case the key-range $[6;11)\subset D$ must be changed to map to $n'_\text{target}$. For this example this is equivalent to merging the old entry $(-\infty;6)$ with $[6;11)$ into $(-\infty;11)$. Now, all nodes in $N_\text{sources}$ can be removed from the index.

	\item \textbf{Split child node horizontally into k nodes:} This is the inverse mutation of M$_3$. 
%
	Figure~\ref{fig:m4} shows an example.


	\item \textbf{Merge sibling nodes vertically:} 
	We set node~$n_\text{parent}:=n$ whose RI maps to at least one other node in $\pi$, if not we abort this mutation.
	From the set of nodes mapped to by $n_\text{parent}$ we randomly select a child node  $n_{\text{child}} \in \text{nodes}(n_\text{parent}$.RI$)$ using the following restriction: $n_{\text{child}}.p = n_{\text{parent}}.p$.
	To merge $n_{\text{child}}$ into $n_{\text{parent}}$, we then need to perform the following updates: First we need to move all data in $n_{\text{child}}.\text{DT}$ to the parent node:
	\[n_{\text{parent}}.\text{DT} = n_{\text{parent}}.\text{DT} \cup n_{\text{child}}.\text{DT}\]
	In the following we need to move potential child nodes $n'$ of $n_{\text{child}}$ to the parent node $n_{\text{parent}}$: 	
	\[\forall_{d \in D_{\text{parent}}}\; n_{\text{child}} \in n_{\text{parent}}.\text{RI}(d)\] 
	\[\Rightarrow n_{\text{parent}}.\text{RI}(d) = n_{\text{parent}}.\text{RI}(d) \setminus \{n_{\text{child}}\} \cup n_{\text{child}}.\text{RI}(d)\]
	where $D_{\text{parent}}$ is the domain of $n_{\text{parent}}.\text{RI}$.
	In other words: We remove all mappings to the merged node $n_{\text{child}}$ and replace them with mappings to the child nodes of $n_{\text{child}}$.
	For our actual implementation, we restrict ourselves to the merge of a single parent-child-pair during a single mutation. 
	Merge operations for longer chains of nodes can easily be achieved by recursively executing the merge operation on the same node.
	Figure~\ref{fig:m5} shows an example: We select the root node as $n_{\text{parent}}$ and its left child node as $n_{\text{child}}$ which we want to merge into the root node. To achieve this we first merge all data contained in $n_{\text{child}}$.DT into $n_{\text{parent}}$.DT.
	In $n_\text{parent}$.RI, we need to remove the mapping to $n_\text{child}$ and replace them with mappings to the children of $n_\text{child}$. In this case, we remove the key-range $(-\infty;6)\subset D$ and replace it with the corresponding entries of $n_\text{child}$.RI. For this example this is equivalent to inserting the entries $(-\infty;2)$ and $[2;6)$ into $n_{\text{parent}}$.RI.

	\item \textbf{Split child node vertically into k nodes:} This is the inverse operation of M$_5$.
	Figure~\ref{fig:m6} shows an example.
\end{enumerate}

%
%
%

\vspace*{-0.3cm}

\subsection{Fitness Function}
\label{subsec:Fitness}


The fitness function is used to measure the performance of a single physical index and describes what to optimize by the genetic algorithm (either by minimizing or maximizing its value).
Its definition can be chosen freely depending on the optimization goal.
We have chosen to optimize our index structures for the runtime given a specific workload consisting of point and range queries.
We therefore define the fitness function $f$: Physical Index $\times$ Workload$ \rightarrow \varmathbb{R}$ to be minimized in the following way:
$f(\pi, W) = r(\pi, W)_c$.
$\pi$ denotes the physical index (the individual) to evaluate,  $W$~is a sequence of queries and denotes the workload of the specific experiment. 
$r(\pi, W)_c$ is the median runtime measured for this physical index on the workload over $c$ runs. 
The fitness function can also easily be adapted to factor in other optimization goals like memory- or energy-efficiency.
Other interesting extensions include regularization, i.e.~index complexity could be punished (similar to model complexity in ML). 
Furthermore we could punish or incentivize the filling grade of leaves, e.g.~if leaves are fully packed, this is beneficial for read-optimized indexes but for inserts can quickly lead to structural modifications of the tree.
However, if leaves are only partially filled, many inserts can be handled by leaf-local changes.
All these requirements can be modeled into the fitness function.

%
%
%
%

%
%
%
%
%
%
%
%
%
%

\vspace*{-0.2cm}

\section{Related Work}
\label{sec:rw}

\noindent\textbf{Handcrafted Indexes.} Since the original B-tree-paper~\cite{DBLP:journals/acta/BayerM72} in 1972, B-trees have become a workhorse in database systems.
Since then a myriad of B-tree-variants and -improvements have been proposed~\cite{DBLP:conf/vldb/RaoR99,DBLP:conf/sigmod/RaoR00,DBLP:conf/damon/SchlegelGL09,DBLP:conf/sigmod/KimCSSNKLBD10}.
Other classes of handcrafted index structures include radix-trees like Judy-arrays~\cite{judy} and its modern SIMDified incarnation ARTful~\cite{DBLP:conf/icde/LeisK013}. Moreover, considerable work has been done in the past years to better understand the performance of hash tables which are widely used in query processing~\cite{DBLP:conf/icde/Alvarez0CD15,DBLP:journals/pvldb/0007AD15}.
    

\noindent\textbf{Learned Indexes.} The core task of a learned index~\cite{DBLP:conf/sigmod/KraskaBCDP18} is to provide an index on a densely packed, sorted array.
The main idea is to manually define an (outer) B-tree-like structure, typically a two-level tree (coined RMI by the authors).
Then, inside each node, rather than performing a binary search on the keys contained in that node --- as done in a textbook B-tree --- a learned regression function
is used to predict the position in the sorted array.
Care has to be taken to avoid prediction errors.
This is done through an error correction method: the prediction actually defines a range which must be post-filtered through a different algorithm like binary or interpolation search. 
The biggest advantage of a `learned index' is that no space is required to store pivots in internal nodes thus allowing for high branching factors.
Like our work, the original work was a read-only index.
It bulkloaded the index top-down, but as with any other B-tree like structure, bottom-up bulkloading up is also possible~\cite{DBLP:conf/sigmod/KipfMRSKK020} and actually easier.
Later on different proposals were made to use different regression techniques~\cite{DBLP:journals/corr/abs-1911-13014} and support inserts and deletes~\cite{DBLP:conf/sigmod/DingMYWDLZCGKLK20,DBLP:journals/pvldb/FerraginaV20}. Also note that the RMIs make a couple of other assumptions that may not always hold in practice~\cite{DBLP:conf/cidr/Crotty21}.
As illustrated in Figure~\ref{fig:logicalindex:example3} already, an RMI is just one special configuration in GENE: an RMI is (1)~a logical index: classical B-tree (however, fixed number of layers, balancing enforced, high fan-out), (2)~a physical index: node internal search constrained to use some form of linear regression.
In other words, an RMI handcrafts its logical structure.
Then, inside its nodes it uses a fixed physical regression method to learn a CDF.
In contrast, we allow for optimizing the structure \textit{and} the search methods and data layouts used inside nodes.
Thus, we fully embrace the orthogonality of \textit{learning a model only inside} a node vs \textit{optimizing the entire index structure}.
Our approach aims at optimizing the entire index structure not only learning weights in a handcrafted structure.

\noindent\textbf{Periodic Tables and Data Calculator.} The work by Stratos Idreos et.al.~on semi-automatic data structure design is truly inspiring.
In their vision paper~\cite{DBLP:journals/debu/IdreosZADHKGMQW18} they aim at a complete dissection and classification of the individual primitives used to design data structures.
They sketch the huge design space of indexes and conclude that many quadrants in that space are still unexplored.
They also phrase the high-level vision to synthesize an index from a declarative specification.
Their main idea is to use a fine-grained learned cost models to be able to cost the physical individual index primitives (like scans, binary search, etc.).
However, they go not further to show how this can be achieved concretely. In addition, no split into logical and physical indexes is given which is the key enabler in our approach.
The follow-up work~\cite{DBLP:journals/debu/IdreosZCQWHKDGK19} is another vision paper which goes into somewhat more detail in describing the problem space of this endeavor and proposing a workbench like ```Data Alchemist' architecture'' which is a semi-automatic design tool.
However, again no experiments and/or results are shown.
Then, \cite{DBLP:conf/sigmod/IdreosZHKG18} explores a large set of physical index design primitives, benchmarks them,
and uses the results to learn cost models for physical primitives.
This is used to build synthesized cost models for the expected cost of a combination of those physical primitives.
The authors show several indexes where these cost estimates match the actual runtimes very well.
At the same time the paper emphasizes that many physical design primitives and their cost models are missing including compression, concurrency, updates, etc.
In their most recent work~\cite{DBLP:conf/cidr/IdreosDQAHRLJGL19}, they present the concept of design continuums, which unify different data structure designs by introducing common parameters, rules, and domains necessary to describe the underlying individuals.
Using this design continuum, they show how to transition between known data structures, exposing also hybrid designs, and how to extend the continuum by new designs.
Their focus lies on the semi-automated construction of these design continuums which are supposed to support researchers and engineers in finding a close to optimal data structure for a given problem composed of workload and hardware by using it as an inference engine.

There are four important differences to our work: we focus on (1)~fully automatic index structure construction, (2)~we provide a clear separation into logical and physical index components, (3)~we believe that the index design space is simply too big for a practical system to be comprehensively modeled by (learned) cost models one reason being that costs models of different physical primitives are often non-additive and hence not usable for an optimization process. (4)~Optimization time is important but not as critical as in standard query optimization: recall that the creation of an index structure is an offline process (in contrast to the creation of an index instance at query time!). And therefore, it makes a lot of sense to define fitness via actual observed runtime measurements rather than cost models whenever possible.



\noindent\textbf{Generic Frameworks.} A couple of generic indexing frameworks have been proposed in the past, most notably GIST~\cite{DBLP:conf/vldb/HellersteinNP95} and XXL~\cite{DBLP:conf/vldb/SeegerBBKSDS01}.
Those frameworks also aimed at generalizing presumably different index structures into a common software framework.
This in turn allowed architects to implement important database algorithms for the generic index.
The specialized indexes could then relatively easily be adapted  to use the generic algorithms. Prominent examples include generic bulkloading \cite{DBLP:conf/vldb/BerckenSW97} and concurrency control \cite{DBLP:conf/sigmod/KornackerMH97}.
Though that work was inspiring to us, we stress that in our paper we argue on a conceptual level rather than an object-oriented-level.
Moreover, we are primarily inspired by the analogue separation into logical and relational operators without immediately specifying how physical operators get implemented (ONC, vectorization, SIMD, whatever) or even how software interfaces need to be defined, as that is a tertiary concern.
    
\noindent\textbf{DQO.} Recently,  we proposed Deep Query Optimization~\cite{DBLP:conf/cidr/DittrichN20}.
The core idea is to break operators into smaller components which can then possibly be optimized using traditional query optimization technique.
This paper is another inspiration of our work.
However, that work does not go into any detail on how such an idea can be realized in the context of indexing.
It neither details how traditional operators can be split nor how this can be turned into an optimization problem for automatic index creation. 
We fill that gap.

\noindent\textbf{Index Selection.} Index Selection~\cite{10.1145/800184.810505,DBLP:journals/pvldb/KossmannHJS20} operates on a completely different level as our approach.
Instead of coming up with a concrete index structure, in index selection the goal is to determine a suitable set of attributes to index in order to improve the runtime of a workload.
In contrast, in our work we consider how to devise efficient index structures in the first place --- which could then be leveraged in index selection algorithms.

\noindent\textbf{Adaptive Indexing.} As index selection is NP-hard, an interesting strategy is to not consider indexing a binary decision but rather allow indexes to become more and more fine-grained over time.
That is at the heart of adaptive indexing~\cite{DBLP:conf/cidr/IdreosKM07}. 
Several interesting proposals have been made in this space, see~\cite{DBLP:journals/pvldb/SchuhknechtJD13} for a survey.
However, all these indexes are still handcrafted indexes.
In future work, we are planning to revisit some of these techniques, as the DT-field of our logical nodes can be used to mimic many of those techniques.

\noindent\textbf{Genetic Algorithms.} Genetic algorithms are a long known search method for an infeasible search space and have been used in our database community for decades.
Early work by Bennett et al.~\cite{DBLP:conf/icga/BennettFI91} applied a genetic algorithm to search for efficient plans in a query optimizer.
Other papers used similar approaches to improve database testing~\cite{DBLP:conf/vldb/BatiGHS07} or to perform index selection~\cite{DBLP:conf/icaisc/KorytkowskiGNS04, DBLP:conf/seke/NeuhausCWRM19, DBLP:conf/glakes/FotouhiG89}.
We are however not aware of papers tackling the problem of index creation using a genetic algorithm and therefore try to further extend the application area of these algorithms.

\noindent\textbf{Decoupling Logical and Physical Indexes.} Early work on partitioning schemes was done by Hellerstein et al.~\cite{DBLP:journals/jacm/HellersteinKMPS02}. They represent data as a set of partitions where each partition is then (redundantly) mapped to at least one physical replica. In contrast to our work, they do not consider partitioning trees as in our logical indexes and they also do not further detail how to physically implement each partition.
In the field of structural indexing~\cite{DBLP:conf/edbt/AgterdenbosFCV16, DBLP:conf/icdt/PicalausaFHV14, DBLP:journals/is/FletcherGWGBP09} introduce the idea to co-partition (or cluster) tuples in a relational schema using graph partitioning. These graph partitions can then be exploited to answer structural queries which could be difficult to compute using foreign key indexes only. Their work has a completely different goal: while we strive to create a single physical index, they strive to create a graph partitioning which can then be mapped to suitable existing indexes. 
Extending our logical index partitions to their graph co-partitions could be an interesting future extension to GENE.
The GMAP project by Tsatalos et al.~\cite{DBLP:conf/dexa/TsatalosI94, DBLP:journals/vldb/TsatalosSI96} is another interesting work in the area of physical data independence and index design.
In contrast to their work, we focus on the clear difference between a logical and physical index and not the schema and a physical index. Moreover, we automatically generate efficient index structures, while their work only allows the choice of one concrete physical index.


\vspace*{-0.1cm}

\section{Experimental Evaluation}
\label{sec:eval}

In our experiments, we first determine a suitable set of hyperparameters for our genetic framework. Based on those hyperparameters, we then carefully evaluate GENE. We highlight the cost for training and the ability to automatically reach a certain performance baseline. Finally, we show the capability of GENE to match and even beat the performance of several state-of-the-art index structures.

\noindent\textbf{System.}
All experiments were executed on a machine with an AMD Ryzen Threadripper 1900X 8-Core processor with 32 GiB memory on Linux.  Our framework and the respective experiments are implemented in \texttt{C++} and compiled with Clang 8.0.1, -O3.
All experiments are run single-threaded and in main-memory.

\noindent\textbf{Datasets.}
We use three types of datasets.  All datasets consist of unique 64-bit uint keys and a 64-bit payload. In the following, we refer to the keys as \emph{data.keys}. The payload represents the offset of the corresponding key into a sorted array. Therefore, we refer to the payload as \emph{data.offset}. The datasets exhibit a variety of different characteristics like \emph{distribution}, \emph{density}, \emph{domain}, and \emph{size}.
The first dataset \emph{uni$_{\text{dense}}$} contains keys that are uniformly distributed in a dense domain. Concretely, $uni_{\text{dense}}$ contains keys in the range [0, n) where $n$ is the size of the dataset.  The other two datasets, \emph{books} and \emph{osm}, represent real-world datasets with complex distributions and are taken from~\cite{DBLP:journals/corr/abs-1911-13014}. The datasets are sampled-down to our specific data size by uniformly drawing elements without duplicates.
We have two main dataset sizes $100$K and $100$M, depending on the concrete experiment. \autoref{tab:datasets} gives an overview of the datasets.

\begin{table}[h]
\begin{footnotesize}
\begin{tabular}{|c||m{2cm}|p{4cm}|}\hline
\textbf{Dataset} & \textbf{CDF} & \textbf{Properties} \\\hline\hline
    uni$_{\text{dense}}$ & \includegraphics[scale=0.16]{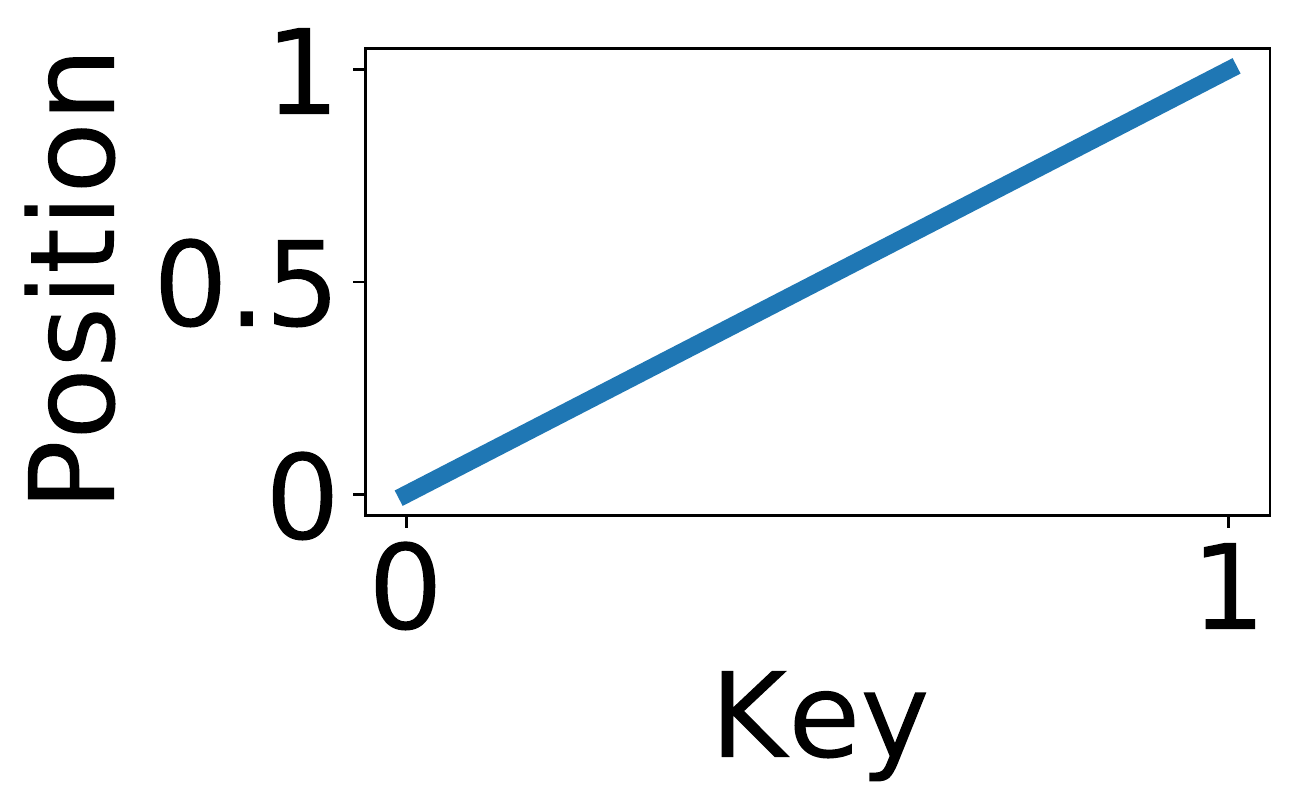} &
    \makecell{
        $n :=$ \# elements (100K, 100M)\\
        64-bit unique unsigned integers
    } \\\hline
    books & \includegraphics[scale=0.16]{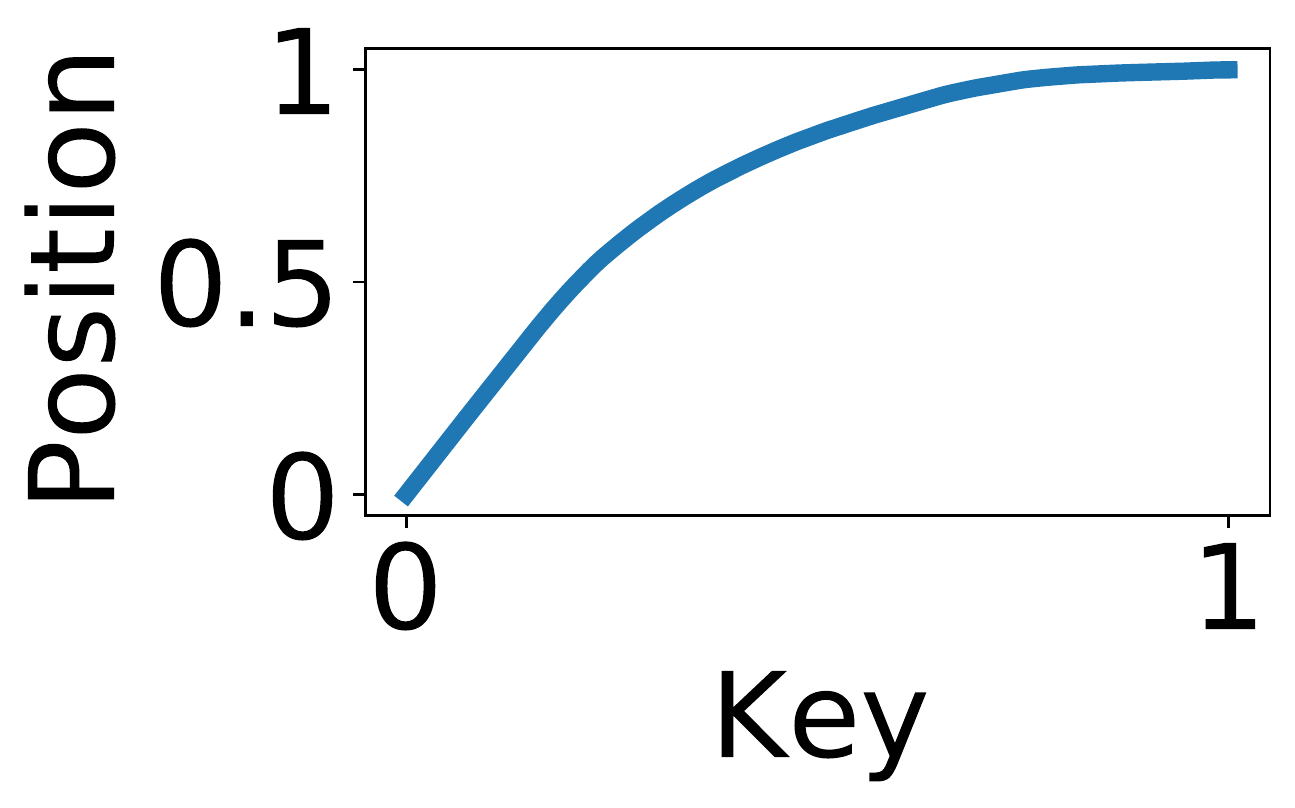} &
    \makecell{
        $n :=$ \# elements (100K, 1M, 10M, 100M)\\
        64-bit unique unsigned integer\\
        Dataset taken from~\cite{DBLP:journals/corr/abs-1911-13014}
    } \\\hline
    osm & \includegraphics[scale=0.16]{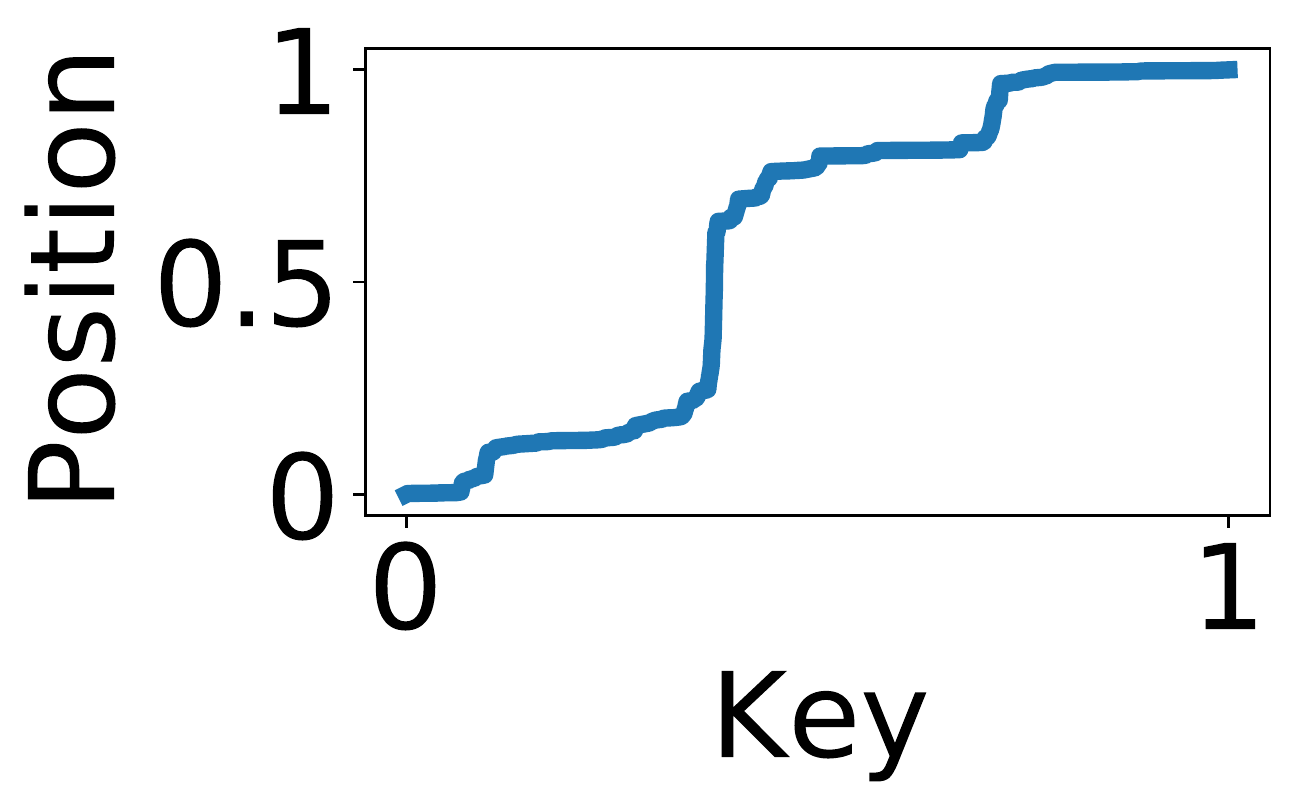} &
    \makecell{
        $n :=$ \# elements (100K, 100M)\\
        64-bit unique unsigned integers \\
        Dataset taken from~\cite{DBLP:journals/corr/abs-1911-13014}
    } \\\hline
    \end{tabular}
\end{footnotesize}
    \caption{Datasets used in the experiments.}
\label{tab:datasets}
\vspace*{-0.7cm}
\end{table}

\noindent\textbf{Workloads.}
We use three classes of workloads: point, range, and mixed point and range query workloads.  For the moment, all our workloads are read-only, i.e.~we do not consider \emph{insert}, \emph{delete}, or \emph{update} statements. Note however, that our generic framework still supports insertions and deletions. In addition, \emph{update} statements would not alter the structure of the index so we could easily integrate them into our framework.
\autoref{tab:workloads} summarizes the basic workload types.  Point(data, idx$_{\text{min}}$, idx$_{\text{max}}$) represents a point query workload where the keys to lookup are taken from the keys in the dataset \emph{data} by selecting indices in the subdomain [idx$_{\text{min}}$, idx$_{\text{max}}$) $\subseteq [0, n)$ with a uniform distribution. Likewise, Range$_{sel}$(data, idx$_{\text{min}}$, idx$_{\text{max}}$) describes a range query workload consisting of pairs specifying the lower bound and upper bound of the query. The lower bound is drawn with a uniform distribution in the index domain [idx$_{\text{min}}$, idx$_{\text{max}}$ - data.size * sel) $\subseteq [0, n)$ and the upper bound is set based on the dataset size and the given selectivity \emph{sel}.  If the domain is not explicitly specified, we assume it to cover the whole dataset. Mix(data, $P$, $R$) represents a mix of point and range queries with $P$ and $R$ being sets of point and range query workloads, respectively, based on \emph{data}.
Note, that in contrast to the datasets, our workloads may contain duplicates.

\begin{table}[h]
\resizebox{\linewidth}{!}{
\begin{tabular}{|c||m{3cm}|m{5.4cm}|}\hline
\textbf{Workload} & \textbf{Characteristics} & \textbf{Parameters} \\\hline\hline
    Point(data, idx$_{\text{min}}$, idx$_{\text{max}}$) & point queries in index domain [idx$_{\text{min}}$, idx$_{\text{max}}$) with uniform distribution & \makecell[l]{[idx$_{\text{min}}$, idx$_{\text{max}}$) $\subseteq [0, n)$} \\\hline

    Range$_{sel}$(data, idx$_{\text{min}}$, idx$_{\text{max}}$) & range queries in index domain [idx$_{\text{min}}$, idx$_{\text{max}}$) with uniform distribution and selectivity $sel$ & \makecell[l]{[idx$_{\text{min}}$, idx$_{\text{max}}$) $\subseteq [0, n)$\\sel $\in [0, 1]$}\\\hline

    Mix(data, $P$, $R$) & mix of point and range query workloads with $P$ and $R$ being sets of respective workloads based on \emph{data} & \makecell[l]{$P$ $:= \{p | p \text{ is Point}\text{(data, idx$_{\text{min}}$, idx$_{\text{max}}$)}\}$ \\ $R$ $:= \{r | r \text{ is Range}_{sel}\text{(data, idx$_{\text{min}}$, idx$_{\text{max}}$)}\}$} \\\hline
\end{tabular}
}
    \caption{Workloads used in the experiments.
    }
\label{tab:workloads}
\vspace*{-0.7cm}
\end{table}

As already showcased in Sections~\ref{sec:generic-indexing:physical}~and~\ref{sec:genetic-index-breeding}, there is a huge search space in designing physical index structures.
Consequently, in our experiments, we focus on the most important data layouts and search algorithms. 
We use the data layouts depicted in \autoref{tab:data-layouts}. 
As search algorithms, we use \textbf{scan}, \textbf{binS}, \textbf{intS}, \textbf{expS}, and \textbf{hashS} described in more detail in \Cref{gindexing:algorithms}.
\begin{table}[h]
\begin{footnotesize}
\resizebox{\linewidth}{!}{
    \begin{tabular}{|c||m{3cm}|m{3cm}|}\hline
    \textbf{Data Layout} & \textbf{Characteristics} & \textbf{Implementation Detail} \\\hline\hline
    \textbf{sorted\_col} & RI and DT have columnar layout for both keys and values. Sorted according to keys. & \texttt{C++} standard library container \texttt{std::vector<Key>} and \texttt{std::vector<Value>} \\\hline
        \textbf{hash} & DT represents hash table mapping keys to their values. RI empty. & \texttt{C++} standard library container \texttt{std::unordered\_map<Key, Value>} \\\hline
        \textbf{tree} & RI and DT represent tree data structure mapping keys to their values. Sorted according to keys. & \texttt{C++} standard library container \texttt{std::map<Key, Value>} \\\hline
\end{tabular}
}
\end{footnotesize}
\caption{Data layouts used in the experiments.}
\label{tab:data-layouts}
\vspace*{-0.65cm}
\end{table}

\subsection{Hyperparameter Tuning} \label{exp:hyperparameters}


\newcommand{\subfigheight}{2.1cm}
\begin{figure}[t!]
	\vspace*{-0.2cm}
	\centering
	\subfigure[\textbf{PQ, }$\mathbf{uni}_{\mathbf{dense}}$]{
		\includegraphics[trim = 2.5mm 2.5mm 2.5mm 2.5mm, clip, width=0.332\linewidth,height=\subfigheight,keepaspectratio]{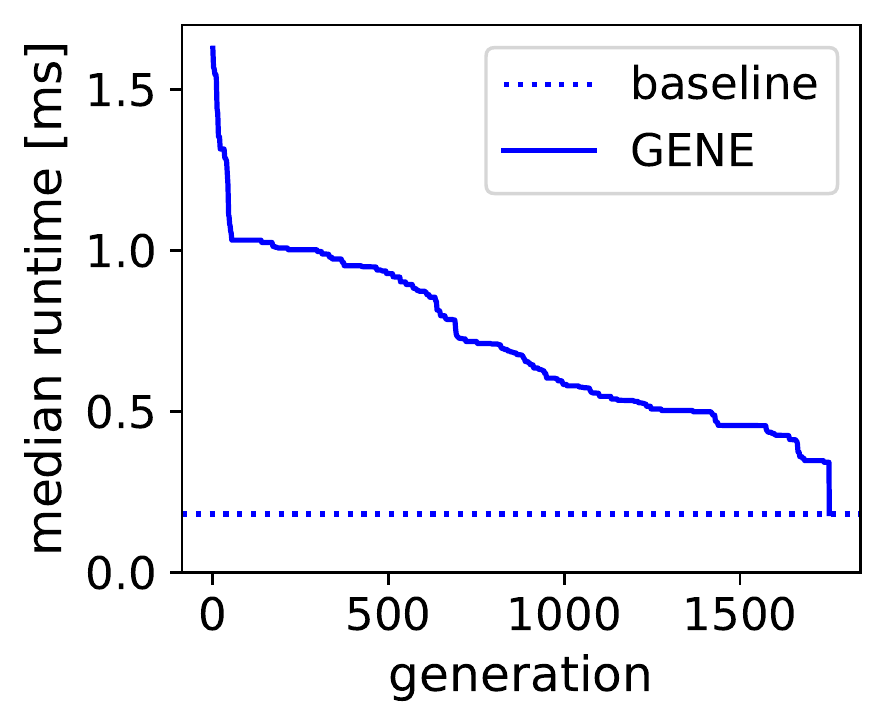}
		\label{fig:baselines_genetic_hashtable}
	}
	\subfigure[\textbf{RQ, }$\mathbf{uni}_{\mathbf{dense}}$]{
		\includegraphics[trim = 8.5mm 2.5mm 2.5mm 2.5mm, clip, width=0.297\linewidth,height=\subfigheight,keepaspectratio]{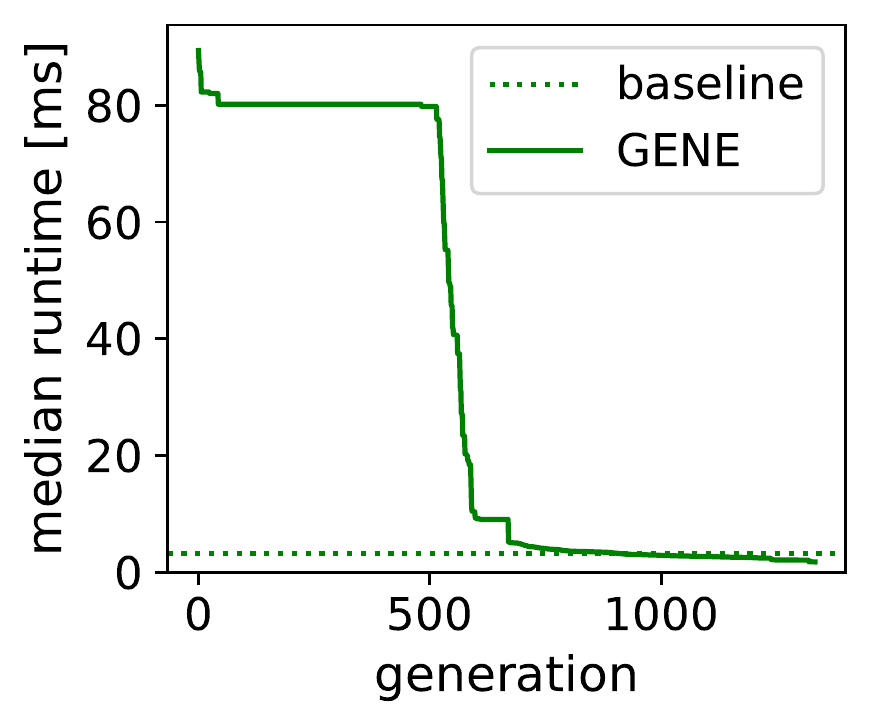}
		\label{fig:baselines_genetic_btree}
	}
	\subfigure[\textbf{Mixed, }$\mathbf{uni}_{\mathbf{dense}}$]{
		\includegraphics[trim = 8.5mm 2.5mm 2.5mm 2.5mm, clip, width=0.283\linewidth,height=\subfigheight,keepaspectratio]{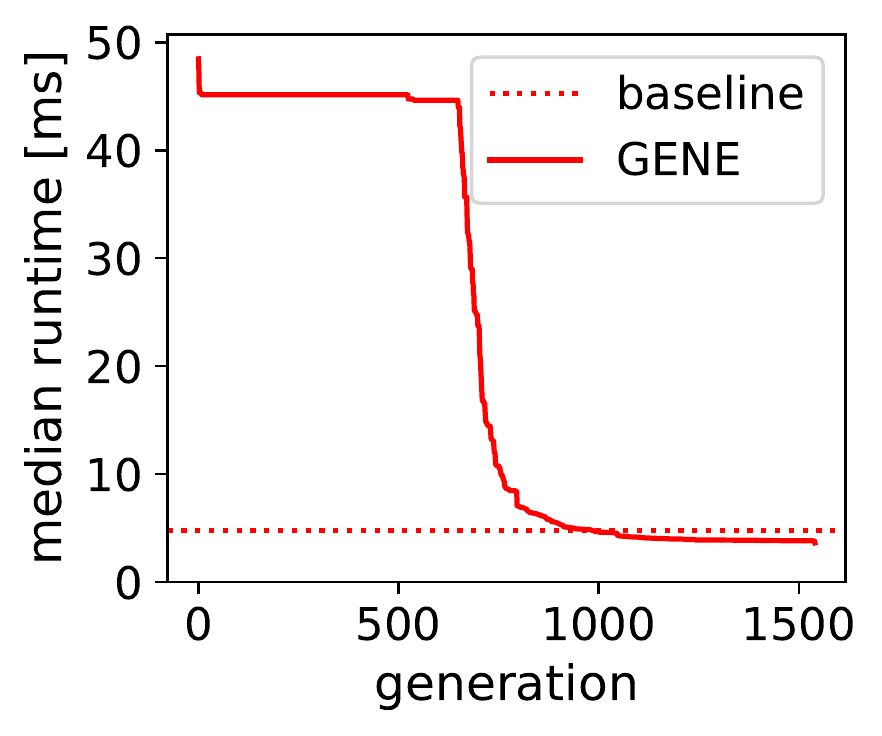}
		\label{fig:baselines_genetic_split}
	}\\
	\subfigure[\textbf{PQ, books}]{
		\includegraphics[trim = 2.5mm 2.5mm 2.5mm 2.5mm, clip, width=0.332\linewidth,height=\subfigheight,keepaspectratio]{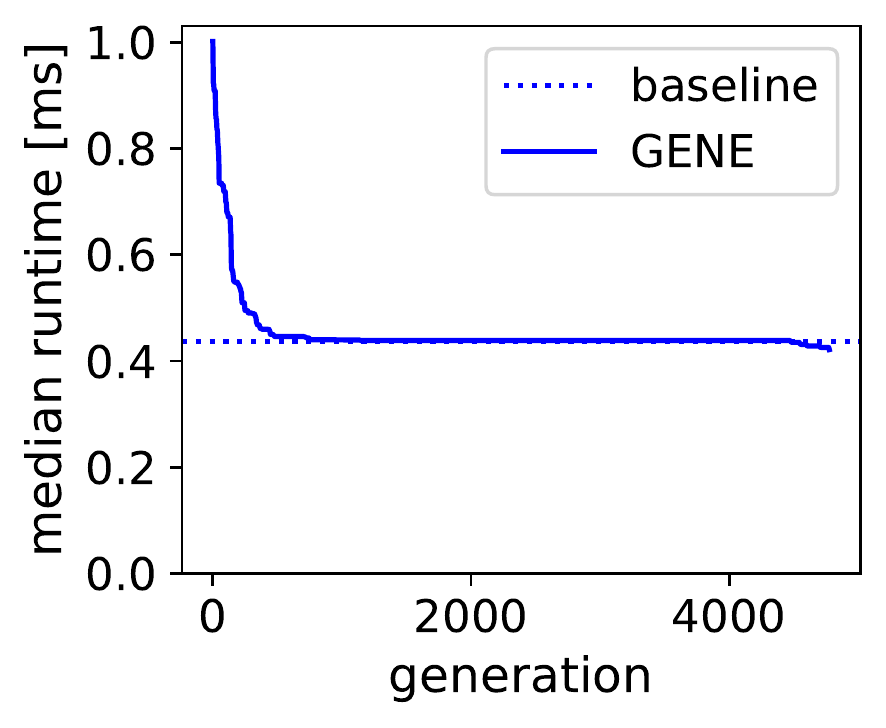}
		\label{fig:baselines_genetic_hashtable_wiki}
	}
	\subfigure[\textbf{RQ, books}]{
		\includegraphics[trim = 8.5mm 2.5mm 2.5mm 2.5mm, clip, width=0.297\linewidth,height=\subfigheight,keepaspectratio]{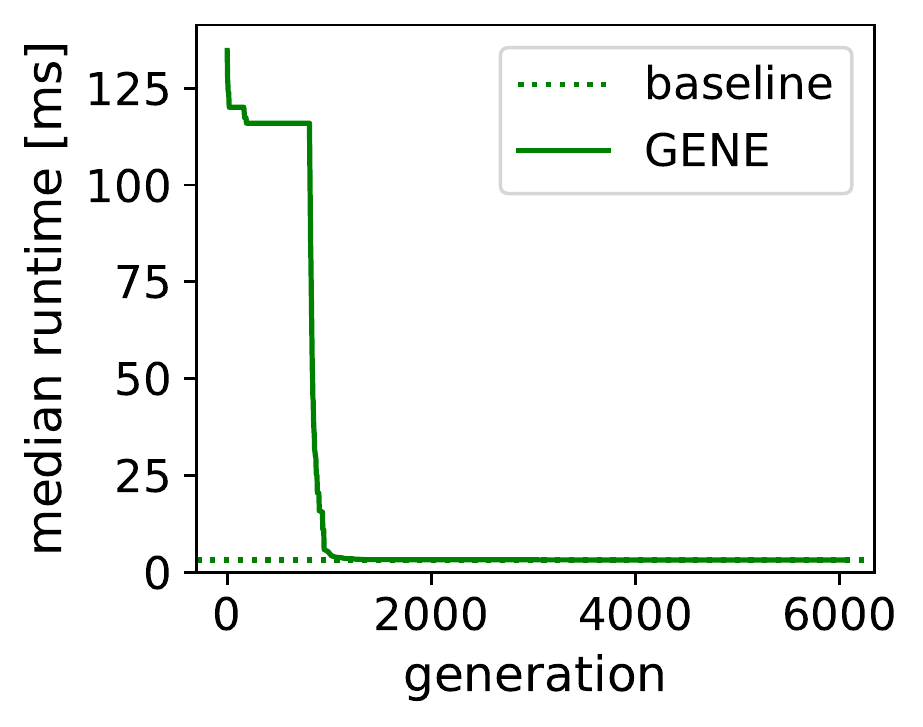}
		\label{fig:baselines_genetic_btree_wiki}
	}
	\subfigure[\textbf{Mixed, books}]{
		\includegraphics[trim = 8.5mm 2.5mm 2.5mm 2.5mm, clip, width=0.283\linewidth, height=\subfigheight, keepaspectratio]{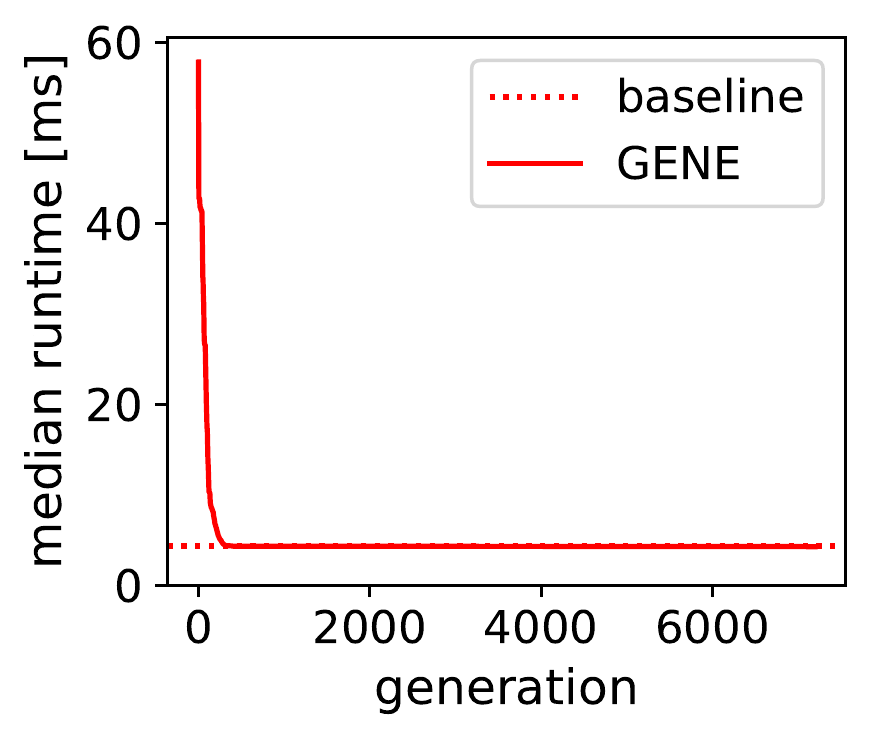}
		\label{fig:baselines_genetic_split_wiki}
	}\\
	\subfigure[\textbf{Upscaling, PQ, books}]{
		\includegraphics[trim = 2.5mm 2.5mm 2.5mm 2.5mm, clip, width=0.332\linewidth,height=\subfigheight,keepaspectratio]{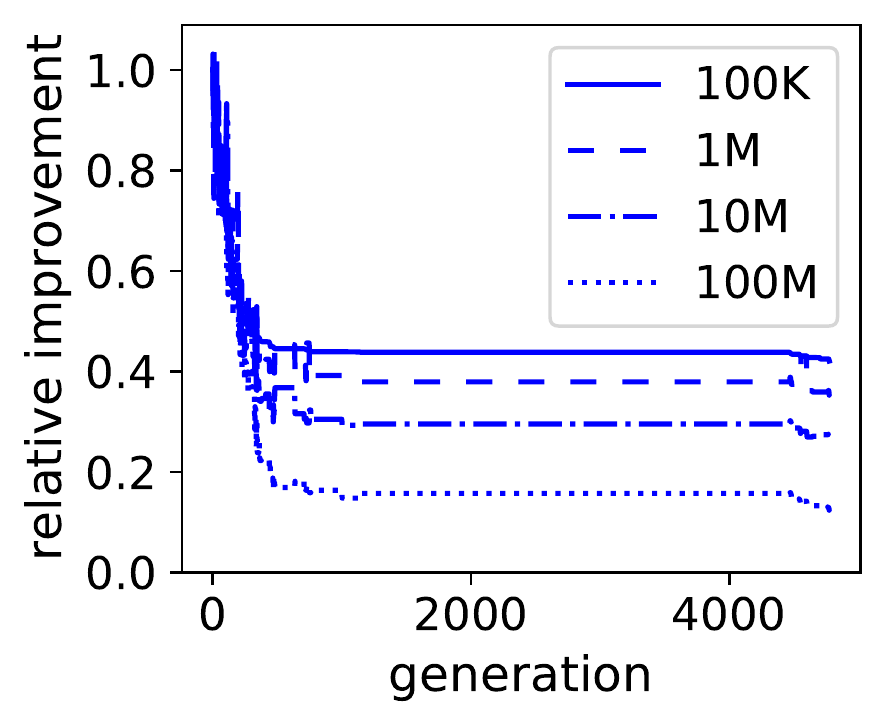}
		\label{fig:baselines_genetic_hashtable_wiki_upscaling}
	}
	\subfigure[\textbf{Upscaling, RQ, books}]{
		\includegraphics[trim = 8.5mm 2.5mm 2.5mm 2.5mm, clip, width=0.297\linewidth,height=\subfigheight,keepaspectratio]{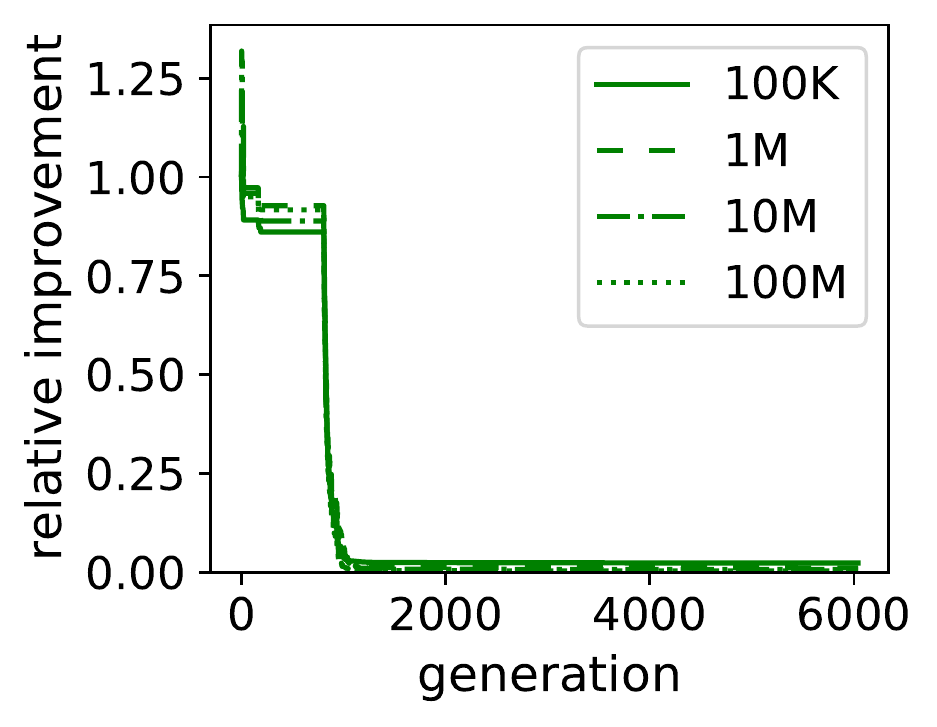}
		\label{fig:baselines_genetic_btree_wiki_upscaling}
	}
	\subfigure[\textbf{Upscaling, Mixed, books}]{
		\includegraphics[trim = 8.5mm 2.5mm 2.5mm 2.5mm, clip, width=0.288\linewidth, height=\subfigheight, keepaspectratio]{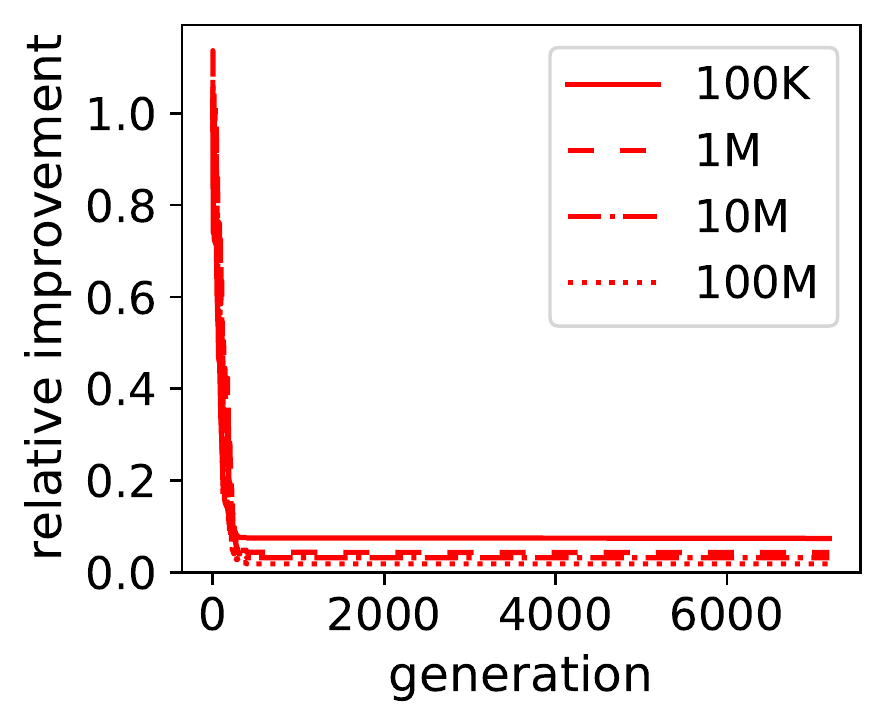}
		\label{fig:baselines_genetic_split_wiki_upscaling}
	}
	\vspace*{-0.4cm}
        \caption{Upper two rows (a-f): GENE approaching handcrafted baselines on three different workloads: A point query only workload (PQ), a range query only workload (RQ) and a mixed workload consisting of 80\% point and 20\% range queries. Bottom row (g-i): Relative improvement compared to the initial index structure after upscaling to dataset sizes of 100K to 100M.}
	\label{fig:baselines_genetic}
	\vspace*{-0.2cm}
\end{figure}
We use a $uni_{\text{dense}}$ dataset of size 100K and vary five different parameters within this experiment:
(1)~number of mutations per generation ($s_{\text{max}}$): $s_{\text{max}} \in \{10, 50\}$,
(2)~maximum population size ($s_\Pi$): $s_\Pi \in \{50, 200, 1000\}$,
(3)~tournament selection size ($s_T$): $s_T \in \{10\%, 50\%, 100\% \text{ of population size}\}$,
(4)~initial population size ($s_{\text{init}}$): $s_{\text{init}} \in \{10, 50\}$,
(5)~population insertion criterion ($q$): Instead of taking the median of the subset drawn during tournament selection, we define a percentile $q$ to be reached for a mutated individual to be inserted into the population: $q \in \{0\%, 50\%, 100\%\}$. For the 0\% percentile, we always insert the mutated individual, for the 100\% percentile we only add it if it is better than the previous best individual within the tournament selection subset.

\begin{table}[h!]
        \vspace{-.1cm}
\resizebox{1.0\linewidth}{!}{
	\begin{footnotesize}
	\begin{tabular}{rrrrrrrr}
		\toprule
		Rank & $s_{\text{max}}$ & $s_\Pi$ & $s_T$ & $s_{\text{init}}$ & $q$ & median runtime [s] & mean runtime [s] \\
		\midrule
		1 & 10 & 200 & 100\% & 50 & 0\% & 13.72 & 91.72 \\
		2 & 10 & 1000 & 50\% & 50 & 50\% & 14.58 & 26.10 \\
		3 & 10 & 1000 & 100\% & 10 & 50\% & 16.71 & 24.94 \\
		4 & 10 & 1000 & 100\% & 50 & 0\% & 16.87 & 94.48 \\
		5 & 10 & 1000 & 50\% & 10 & 50\% & 18.21 & 158.49 \\
		\bottomrule
	\end{tabular}
	\end{footnotesize}
}
		\caption{Best Genetic Search Configurations (over 5 runs)	\label{tab:best-grid-search-configs}}
        \vspace{-.7cm}
\end{table}

Table~\ref{tab:best-grid-search-configs} shows the best configurations  (based on the median of the 5 runs executed per configuration). 
Given a total number of mutations we want to perform, we conclude that it is more beneficial to use a smaller number of mutations per generation combined with a larger number of generations. 
As the population size has a  limited influence, we decided to keep it very small to reduce the overhead to maintain the population.
We therefore used the following default parameters for the experiments in the following sections:
$s_{\text{max}}=10, s_\Pi=50, s_T=25, s_{\text{init}} = 10$ and $q=50\%$.



\subsection{Rediscover Suitable Baseline Indexes}

In this experiment, we will demonstrate that our genetic algorithm is capable of reproducing the performance of various baseline index structures as known from textbooks.
We consider two different datasets: $uni_{\text{dense}}$ and \emph{books} of sizes 100K, 1M, 10M and 100M. 
We combine each of those two datasets with three different workloads containing 10,000 queries each:
 $\text{Point}(\text{uni}_{\text{dense}})$, Range$_{0.001}(\text{uni}_{\text{dense}})$ and a Mix(uni$_\text{dense}$, $P$, $R$) workload, with P := \{Point$(\text{uni}_{\text{dense}})$\} and R := \{Range$_{0.01}(\text{uni}_{\text{dense}})$\} consisting of $80$\% point and $20$\% range queries.
For each workload, we define a baseline within our generic framework of which we believe it has a decent performance:
For the point query only workload, we assume a simple hash table to perform best which is implemented as an index structure with a single node having the hash data layout.
For the range query only and mixed workload, we assume a B-tree-like structure to offer a decent performance. We initialize the tree to have 100 fully filled leaves, each containing 1,000 elements and a fan-out of 10 for the internal nodes. Each node is configured to use the \texttt{sorted\_col} layout and \texttt{binS}.
We configured GENE to allow nodes to contain up to 100,000 key-value-pairs or 100,000 child partitions (potentially leading to solutions consisting of a single node or solutions with one node per element assembled under a single root node).
In the initial population trees were bulkloaded 
with 100 equally filled leaves and a fan-out of 10, but with randomized data layouts and search methods.
Each experiment is conducted for 8000 generations. 
The genetic search was run on the smallest sample size of 100K elements. Each time we found a new, best individual, we checked if the results carry over to the larger datasets, i.e.~we created new index structures using the same routing information and data layouts and search methods as found by GENE (i.e.~using the exact same index structure), but bulkloaded them with the larger dataset, increasing leaf capacities if necessary. We then evaluated them using the exact same workload as used in the genetic search.

Figure~\ref{fig:baselines_genetic} shows the results. Each plot in the first two lines compares the performance of the baseline to the performance of the genetic algorithm where we plotted the best individual of each generation. 
We plot the curves up to the point of the last improvement.

As we can clearly see, GENE rapidly approaches the baseline. This is mostly due to the fact that GENE can rather easily improve by mutating very inefficient nodes in the beginning. After getting close to the baseline, GENE only finds slight improvements, e.g.~by changing search algorithms within nodes, which are hardly visible on the plot.
The index structures found by GENE are very similar to the baselines:
On the $uni_{\text{dense}}$ dataset, GENE always returned a single node index structure. For the point query only workload, it came up with a single \texttt{hash} node containing all entries, i.e.~exactly the baseline we defined beforehand. For the range query only as well as mixed workload, GENE also reduced the index to a single node, but with \texttt{sorted\_col} data layout and \texttt{intS} search method. This difference is due to the fact that range queries can not be executed efficiently on a hash node. This result is reasonable as a uniformly distributed, dense dataset can easily be modeled by an array with a linear model as search method.
Considering the \emph{books} dataset, the point query workload resulted in a tree with 68 nodes in total, 66 of them being leaves. All but one leaf are direct children of the \texttt{sorted\_col} root node, the remaining leaf has a single \texttt{tree} node between itself and the root. With 48 nodes, the vast majority of the leaves has a \texttt{hash} data layout. The remaining leaves are of \texttt{sorted\_col} (15) or \texttt{tree} data layout (3). The dominating search method for non-\texttt{hash} nodes is \texttt{binS}, with only 3 exceptions that use \texttt{expS}. The resulting index structure reminds of a partitioned hash map, indicating that GENE indeed approached the expected baseline.
For the range query workload, we obtained an index with similar size, having 44 nodes with \texttt{sorted\_col} data layout in total, 40 of them being leaves. The index has a height of three with the majority of the leaves (38) situated at depth two and only two leaves being one level below. \texttt{BinS} is again the dominating search method for the leaves, with four nodes using \texttt{intS} and two using \texttt{expS} instead. The resulting index structure reminds of a shallow B-tree, indicating that GENE again approached the expected baseline.
For the final mixed workload, the results are similar to the range only workload. We obtained an index of height three with 41 nodes in total (all with \texttt{sorted\_col} data layout), 35 of them being leaves. The majority of the leaves is at depth two, with three leaves being one level above and one leaf being a level below. The dominating search method is again \texttt{binS}, with only 7 leaves using an \texttt{intS} instead. As for the range query only workload, GENE approached a shallow B-tree like index to match the performance of the baseline.
The last line in Figure~\ref{fig:baselines_genetic} shows the improvements of the scaled index structures for the books dataset. Each line represents the relative improvements compared to the best index structure of GENE's initial population, upscaled to the indicated dataset sizes of 100K (the size on which the search was conducted) up to 100M. We can clearly see that an improvement in the solid line representing the training data nearly always results in a very similar improvement for the upscaled index structures. The overall, relative improvement becomes even bigger with increasing dataset size, indicating that is most likely sufficient to run GENE on a sample of the data to obtain a decent index structure, highly reducing the necessary search time. If best possible performance is the ultimate goal, then GENE can again be applied to the upscaled index structure resulting from the sample to perform further fine tuning.

We also experimented with an additional, mixed workload again consisting of 10,000 queries with a 80\% / 20\% point to range query ratio, based on the $uni_{\text{dense}}$ dataset.
However, this time we chose the queries to be normally distributed around key 75,000 with a standard deviation of 10,000, i.e.~the queries were mainly focused on the upper half of the key domain.
Our GENE algorithm again decided to shrink the initial index structures considerably, however it stopped after 3500 generations returning a tree with 4 levels and 25 nodes in total, 17 of them being leaves.
The nodes containing the upper half of the key domain were again using the \texttt{sorted\_col} layout and either \texttt{intS} or \texttt{binS}.
The total runtimes of GENE heavily depend on the concrete datasets and workloads. The fastest execution for $uni_{\text{dense}}$ with point query only workload took less than 3 minutes until the last improvement was found. The longest run on the same dataset with range query only workload took about 122 minutes. Performing the additional upscaling steps further influenced the runtimes, leading to execution times of up to 30 hours for the \emph{books} dataset in combination with range query only workload.

    \vspace{-0.2cm}
\subsection{Optimized vs Heuristic Indexes}
\label{subsec:poc}
In this section, we will compare the performance of a GENE index with representatives of different prevalent heuristic index types. \autoref{tab:competitors} gives an overview of the different index types and respective representatives.
For the B+tree implementation we use the commonly used TLX baseline implementation by Bingmann~\cite{TLX}. In particular, we use the specialized B+tree template class \texttt{btree\_map} implementing STL's map container.
The ART implementation is taken from the SOSD benchmark~\cite{sosd-vldb} by Marcus et al. and concretely, we use the implementation \texttt{ARTPrimaryLB} that supports lower bound lookups.
PGM~\cite{Ferragina:2020pgm} by Ferragina et al. provides multiple implementations that support a variety of different functionalities like insertion and deletion support or compression to reduce space usage. Since we are only interested in the lookup performance, we use the default \texttt{PGMIndex} implementation.
We purposely exclude hash tables since they do not support range queries efficiently.
\begin{table}[h]
    \vspace{-0.3cm}
\begin{footnotesize}
    \begin{tabular}{|l||m{1.5cm}|m{3.5cm}|}\hline
    \textbf{Type} & \textbf{Index} & \textbf{Details} \\\hline\hline
        Tree & B-tree & TLX \texttt{btree\_map} \cite{TLX}\\\hline
        Radix & ART & SOSD \texttt{ARTPrimaryLB} \cite{sosd-vldb}\\\hline
        Learned & PGM & PGM \texttt{PGMIndex} \cite{Ferragina:2020pgm} \\\hline
    \end{tabular}
\end{footnotesize}
    \caption{Overview of different index types and representatives of each category.}
    \label{tab:competitors}
    \vspace{-0.7cm}
\end{table}

\begin{figure}[t]
    \centering
    \includegraphics[trim = 0mm 0mm 0mm 0mm, clip, keepaspectratio, width=\linewidth]{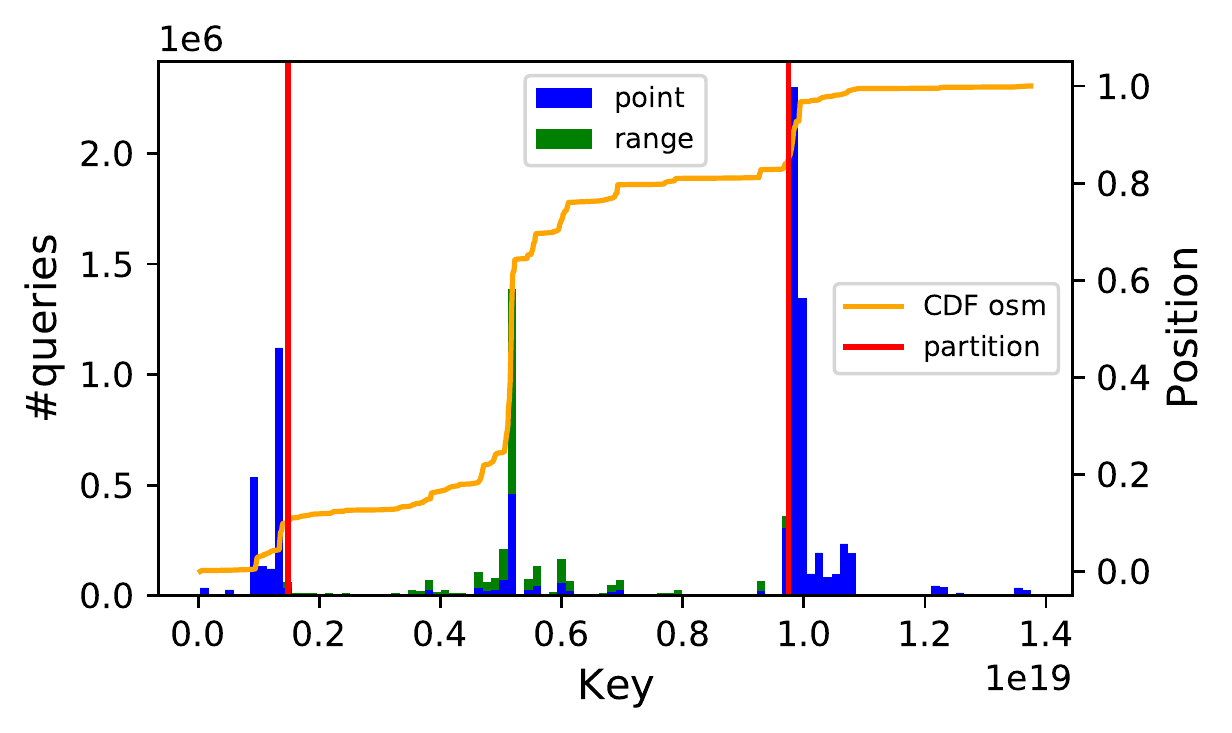}
\vspace*{-0.7cm}
    \caption{Visualization of the experimental setup. The \emph{osm} dataset is shown as CDF while the point and range queries are illustrated as a stacked histogram. The red vertical lines highlight the partition borders.}
    \label{fig:poc_wkl}
\vspace*{-0.3cm}
\end{figure}

We conduct our performance evaluation on the three different datasets, \emph{uni$_{\text{dense}}$}, \emph{books}, and \emph{osm}, each with a size of $n = 100$M data points.
As for the workload, we are going to use a mixed workload consisting of multiple point and range query workloads. Concretely, the workload consists of 1M queries, divided in three point query workloads and one range query workload: Mix(data, $P$, $R$), with $P$ := \{Point(data, 0, 0.1 $\cdot$ n), Point(data, 0.1 $\cdot$ n, 0.85 $\cdot$ n), Point(data, 0.85 $\cdot$ n, n)\} and $R$ := \{Range(data, 0.1 $\cdot$ n, 0.85 $\cdot$ n)\}, where data $\in$ \{uni$_{\text{dense}}$, books, osm\}. With that, the queried key domain is essentially split into three partitions at 10\% and 85\% of the data based on the different workloads. The first partition $[0, 0.1 \cdot n)$ exclusively receives point queries representing 20\% of the total workload size. The second partition $[0.1 \cdot n, 0.85 \cdot n)$ receives a mix of both, 10\% point and 20\% range queries, and the third partition $[0.85 \cdot n, n)$ 50\% point queries.
\autoref{fig:poc_wkl} illustrates the workload based on the \emph{osm} dataset.
Since each data point maps a key to its position in a sorted data array, range queries can be translated to finding the position of the lower bound in the index and subsequently scanning the data array. This scan is independent of the underlying index type and can therefore be neglected. Thus, a range query in our evaluation is equivalent to a lower bound lookup in the index.
\begin{figure}[t]
    \centering
    \includegraphics[trim = 0mm 6mm 0mm 0mm, clip, keepaspectratio, width=\linewidth]{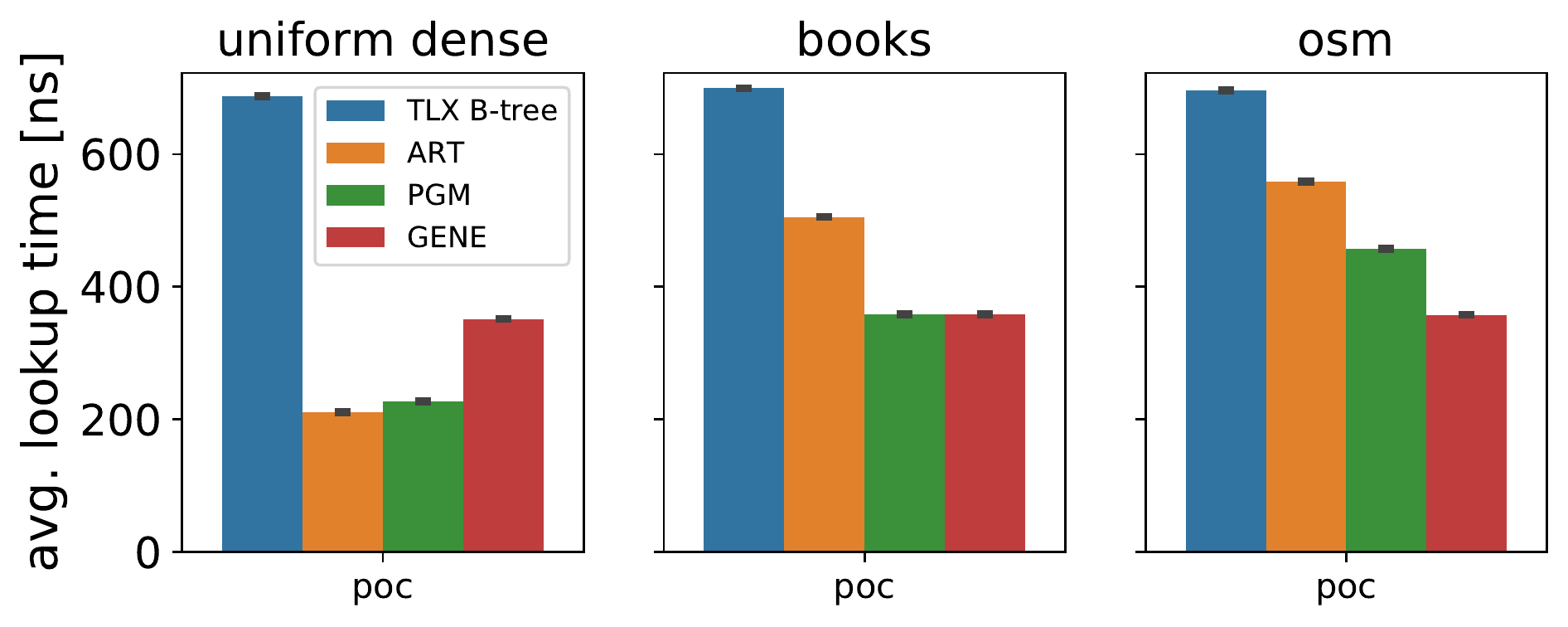}
    \vspace*{-0.5cm}
    \caption{Average index lookup time comparison between three representative state-of-the-art index structures and our GENE index on three different datasets and the corresponding workload described in subsection~\ref{subsec:poc}. The small black bars indicate the standard deviation of five runs, which is negligibly small.}
    \label{fig:poc}
    \vspace*{-0.3cm}
\end{figure}
\begin{figure}[t]
    \centering
    \includegraphics[trim = 0mm 132mm 262mm 0mm, clip, keepaspectratio, width=.95\linewidth, page=2]{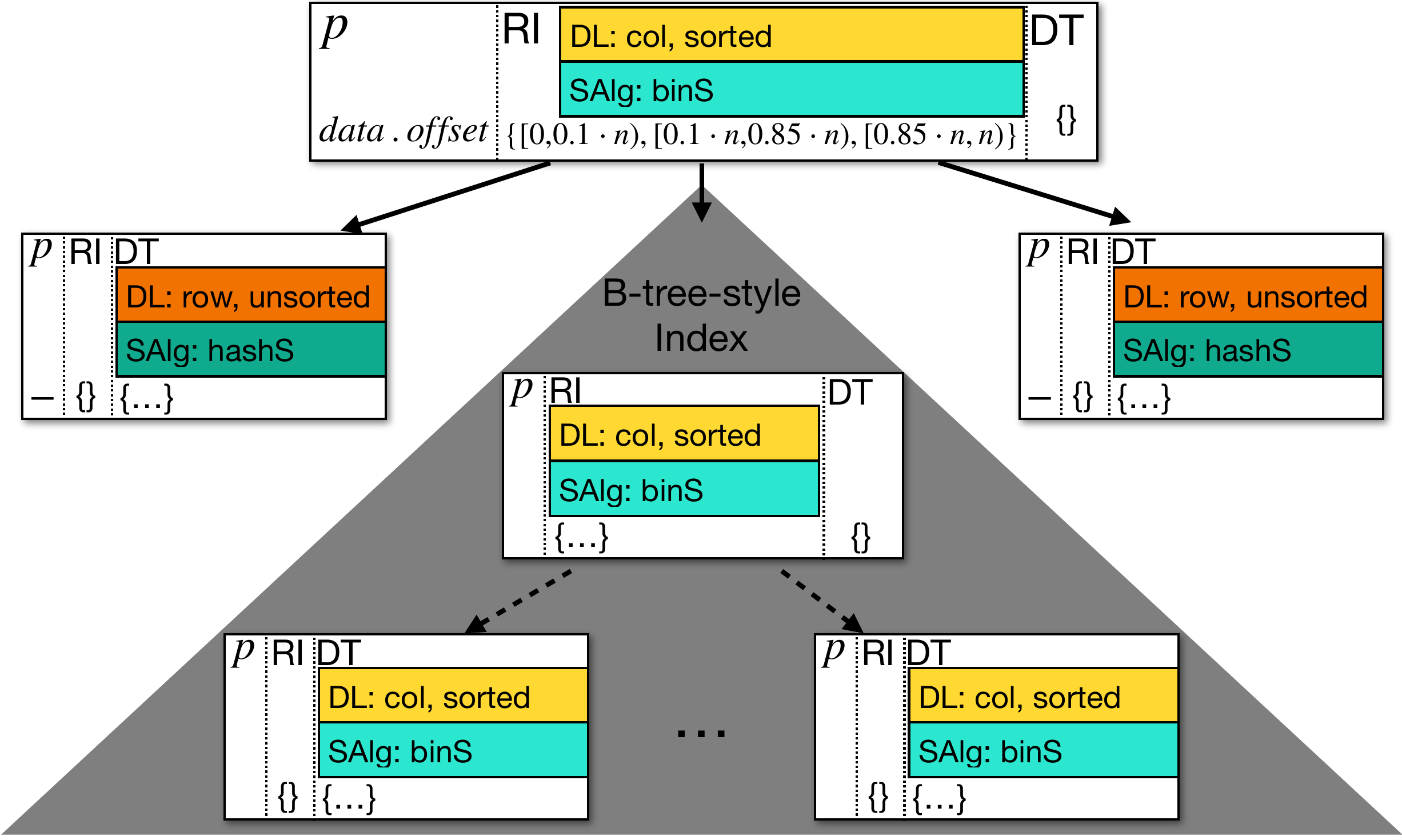}
    \caption{Physical index structure of the GENE index based on the workload partitioning.}
    \label{fig:gene_index}
    \vspace{-0.6cm}
\end{figure}
Our generic implementation allows us to easily replace specific parts of a physical index structure like the data layout or search method. However, this leads to a non-negligible performance overhead mainly due to repeated dynamic dispatches. To be competitive with the other baselines and state-of-the-art index structures, we provide an additional implementation that specifically contains the concrete physical index structures used in this experiment.
\autoref{fig:gene_index} shows the physical structure of our GENE index. Since the workload domain is split into three partitions with two exclusive point query regions, we bulkload our index structure accordingly. The first and third partition are hash nodes while the second partition represents a B-tree-style index. The root is a sorted array using binary search. 
We randomly shuffle the workload before each execution to avoid caching effects.

\autoref{fig:poc} shows the results of the index structures for different datasets. We report the average index lookup time.  Independent of the underlying dataset, the TLX B-tree requires around 700 ns and is not able to compete with the other indexes. On the uniform dense dataset, ART and PGM both achieve a lower lookup time than GENE. However, for both, a uniform dense dataset is close to the optimal use case. For the two real-world skewed and sparse datasets, our GENE index achieves a competitive or even faster lookup time than the other index structures of around 350 ns.

We are well aware that this is a very specific use case, however, it showcases that there are indeed scenarios where an optimized GENE index can outperform a state-of-the-art (heuristic) data structure. Expanding the covered design space by GENE, i.e.~the available data structures and search algorithms, and automatically finding those scenarios is part of future work. In conclusion, our proof of concept emphasizes that there are use cases in which GENE is able to achieve a competitive or even superior performance than state-of-the-art index structures and therefore, confirms its validity.

\section{Conclusion and Future Work}
\label{sec:conclusion}
\label{sec:futurework}

\noindent\textbf{Conclusions.}
This paper has opened the book for automatically generated index structures. We have proposed a powerful generic indexing framework on the logical and physical level analogue to logical and physical operators in query processing and optimization.  We have shown that by clearly separating the logical and physical dimensions of an index, a huge number of existing (physical) indexes can be represented in our generic indexing framework. Furthermore, we introduced \textit{Genetic Generic Generation of Indexes~(GENE)}. Given a workload, GENE can come up with an efficient physical index structure automatically. Our initial experimental results outlines the potential and efficiency of our approach. 


\noindent\textbf{Future Work.} This paper is obviously just a starting point of a much longer story. There are many possible exciting research directions ahead, including:
\begin{enumerate}[itemindent=0.45 cm,labelsep=0.1cm,leftmargin=0cm]
\item code-generation, similar to generating code for the most efficient physical \textit{plan} found, generate code for the most efficient \textit{physical index structure} found,
\item \textit{The Index Farm}: we plan to open source our framework: the goal is that people submit a workload on a web page and the framework emits suitable source code for an index structure,
\item runtime adaptivity: how to mutate structurally, this can also simulate the adaptive indexing family of index structures,
\item updates: simple updates are trivial, i.e.~if a value assigned to an existing key is changed, no structural modifications required at this point);
    the generic framwork already supports inserts and deletes but adapting to insertion and deletion workloads is more complex and beyond the scope of this paper, 
\item scalability: extend our scalability experiments to evaluate workloads only on subtrees affected by mutations 
using cost functions to prioritize expensive partitions when drawing nodes for mutations
\item effects of non-empty DT-fields in internal nodes,
\item extend GENE to support more data layouts, search algorithms, and hardware acceleration (SIMD).
\end{enumerate}

\clearpage

\bibliographystyle{ACM-Reference-Format}
\bibliography{main}

\end{document}